\documentclass[12pt]{article}

\usepackage{amssymb, amsmath}
\usepackage{graphicx, caption}
\usepackage{booktabs}
\usepackage{multirow}
\usepackage{enumerate}
\usepackage{natbib}
\usepackage{hyperref}

\usepackage{pdfpages}

\begin{document}

\def\spacingset#1{\renewcommand{\baselinestretch}%
{#1}\small\normalsize} \spacingset{1}


{
  \title{\bf Bayesian sample size determination using commensurate priors to leverage pre-experimental data}
  \author{
  Haiyan Zheng$^{1,\dagger}$, Thomas Jaki$^{2,3}$, James M. S. Wason$^{1,3}$ \thanks{
    The authors gratefully acknowledge \textit{the UK Medical Research Council (MC\rule{2mm}{0.4pt}UU\rule{2mm}{0.4pt}00002/6, MC\rule{2mm}{0.4pt}UU\rule{2mm}{0.4pt}0002/14). This report is independent research arising in part from Prof Jaki's Senior Research Fellowship (NIHR-SRF-2015-08-001) supported by the National Institute for Health Research. The views expressed in this publication are those of the authors and not necessarily those of the NHS, the National Institute for Health Research or the Department of Health and Social Care (DHSC).}}\hspace{.2cm} \\[8pt]
    \textit{$^1$Population Health Sciences Institute, Newcastle University, U.K.} \\
\textit{$^2$Department of Mathematics and Statistics, Lancaster University, U.K.} \\
\textit{$^3$MRC Biostatistics Unit, University of Cambridge, U.K.} 
\\[8pt]
$^\dagger$Email: \href{mailto:haiyan.zheng@newcastle.ac.uk}{haiyan.zheng@newcastle.ac.uk} 
}
\date{}
  \maketitle



\bigskip
\begin{abstract}
This paper develops Bayesian sample size formulae for experiments comparing two groups. We assume the experimental data will be analysed in the Bayesian framework, where pre-experimental information from multiple sources can be represented into robust priors. In particular, such robust priors account for preliminary belief about the pairwise commensurability between parameters that underpin the historical and new experiments, to permit flexible borrowing of information. 
Averaged over the probability space of the new experimental data, appropriate sample sizes are found according to criteria that control certain aspects of the posterior distribution, such as the coverage probability or length of a defined density region. Our Bayesian methodology can be applied to circumstances where the common variance in the new experiment is known or unknown. Exact solutions are available based on most of the criteria considered for Bayesian sample size determination, while a search procedure is described in cases for which there are no closed-form expressions. We illustrate the application of our Bayesian sample size formulae in the setting of designing a clinical trial. Hypothetical data examples, motivated by a rare-disease trial with elicitation of expert prior opinion, and a comprehensive performance evaluation of the proposed methodology are presented.
\end{abstract}

\noindent%
\vfill

\newpage
\spacingset{1.5} 

\section{Introduction}
\label{sec:Intro}

Suppose that random variables $X_A$ and $X_B$ have probability density functions (pdf) of some known form, denoted by $f(x_j; \mu_j), j = A, B$, where $\mu_j$ are the group-specific parameters, for example, the population means, of inferential interest. We further suppose that the observations are independent and indentically distributed (i.i.d.) with their pdfs $f(x_j; \mu_j)$ conditional on the unknown  $\mu_j$.  Consider the problem of comparing $\mu_A$ and $\mu_B$, both supposed to be one-dimensional parameters for simplicity, based on two samples. A classical statistical problem is how to choose the sample size for the experiment to allow good quality inference.  There have been many sample size determination (SSD) methods in the literature; the main ways in which they vary are whether: (a) the inferential goal is parameter estimation or hypothesis testing, (b) additional parameters relating to the experimental design, e.g., variance of a normal distribution, are assumed known or unknown, and (c) the analysis approach is frequentist or Bayesian.

Conventional SSD \citep{Desu1990} for such experiments has often been carried out to control certain aspects of the sampling distribution of a test statistic suitable for comparing two group means. This is typically considered from a frequentist perspective that operating characteristics, e.g., type I error rate and power, should be specified for falsely and correctly detecting a meaningful magnitude of the difference in means.  For data that are assumed to be i.i.d. normal, SSD may be a function also of nuisance parameters such as unknown variances,  denoted by $\sigma_j, \, j = A, B$. It is not uncommon that investigators specifiy a fixed value for the unknown $\sigma_j$, which could deviate far from the true value.
Consequently, this may leave the SSD inaccurate, or only locally optimal, due to the dependence on arbitrarily guesses.
Formulating the problem in the Bayesian framework has been argued to be more advantageous, since it allows uncertainty to be described in a prior for the nuisance parameters \citep{O'Hagan2004}.  
Moreover, it brings about the possibility of incorporating pre-experimental information, if available, in a prior for the nuisance parameters and/or the parameter(s) of inferential interest. For these reasons, considerable attention has been given to Bayesian SSD; see, for example, \citet{JRSSD:Adcock1997, AS:Clarke2006}.

This paper considers robust Bayesian SSD approaches for experiments basing inference on two samples, for which we suppose pre-experimental data can be used to specify a prior for the parameter(s).  
Two main kinds of methodology written in the literature are `hybrid classical and Bayesian' and `proper Bayesian' SSD \citep{Spiegelhalter2004}. 
By `hybrid classical and Bayesian', it indicates that  the final analysis of data would be conducted in the classical, frequentist framework despite the use of a prior to account for uncertainty \citep{SiM:Spiegelhalter1986, SS:Lee2000}.
An adequate sample size would thus be chosen to ensure that the predictive power, obtained by averaging the frequentist power function over the prior distribution for the unknown parameter(s), reaches a desired target level. 
A variant to this type is the `two-prior approach' \citep{MDM:O'Hagan2001, SS:Wang2002, JRSSA:Sahu2006}, so-called because distinct priors are specified for the SSD and analysis of data, respectively.  \citet{JRSSA:DeSantis2007} considers using power priors \citep{SS:Ibrahim2000, SIM:Neuenschwander2009}, which raise the likelihood of pre-experimental data to a power between 0 and 1, as the {\em design prior}, along with a non-informative {\em analysis prior} to yield posterior quantities free of the impact from the pre-experimental data.  \citet{SiM:Brutti2009} use a {\em design prior} that harmonises the sampling distribution of the new experimental data, and accommodate divergent pre-experimental information in a mixture of priors conjugate to the normal likelihood as the {\em analysis prior}.

By contrast, `proper Bayesian' SSD approaches refer to those using the same prior for both design and analysis of the new experiment. \citet{JRSSD:Joseph1995} derive formulae for binomial experiments comparing two proportions; specifically, sample sizes are sought to ensure, for example, a desired level of coverage probability or width of a defined interval of the posterior `success' rate. 
\citet{JRSSD:Joseph1997} concentrate on normal distributions and use normal-gamma conjugate priors, for experiments that estimate either single normal means or the difference between two normal means.
In the context of clinical trials in medical research, \citet{SiM:Whitehead2008} develop Bayesian methods resembling frequentist formulations of the SSD problem in exploratory trials: the effect size $\mu_\Delta = \mu_A - \mu_B$ is tested based on posterior interval probabilities that mimic the frequentist error rates.  These fully Bayesian approaches shed light on the option of incorporating pre-experimental data into a prior for both design and analysis consistently, without undermining the data validity and integrity of the new trial \citep{EMA:E6R2}. The robust Bayesian SSD approach that we will propose in this paper falls within this category.

Our research is partly motivated by the efficient design and analysis of clinical trials that evaluate a new treatment for rare diseases \citep{EMA:smallp2006}. 
It is often infeasible to ask for a sample size achieving the frequentist power, nor to draw an inference solely based on the scant trial data. Pre-trial information, collected from historical studies which had been conducted under similar circumstances, or elicited from expert opinion, could play an essential role. 
Most existing proper Bayesian SSD approaches, as those mentioned above, employ a conjugate prior for algebraic convenience. Under such a framework, however, it is not straightforward to leverage pre-trial information, especially if it has been available from multiple sources. In this paper, we propose basing the SSD on a robust Bayesian model with commensurate priors: we specify predictive priors with respect to the sources of pre-trial information, and parameterise the commensurability between a historical and the new trial parameters explicitly \citep{BIOM:Hobbs2011, BA:Hobbs2012, BIOS:ZhengWason2019}. By placing a Gamma mixture prior on each commensurate parameter, our approach can overcome the notorious prior-data conflict and maintain an appropriate borrowing from each relevant source.   The proposed methodology is generic: it can be applied to areas where there is a need to use pre-experimental data formally through the mechanism of specifying priors. For instance, the sample size for environmental water quality evaluation could often be limited, for which borrowing strength from historical water monitoring data has been considered helpful \citep{Envr:Duan2006}. Other examples include quality control in management and engineering \citep{MR:Kleyner1997, JAS:Stamey2006, IJRS:Khorate2018}.

The remainder of this paper is structured as follows. 
In Section \ref{sec:methods}, we develop a robust Bayesian model for leveraging pre-experimental information from multiple sources, where customised borrowing by source of information is possible. Focusing on normally i.i.d. data, Bayesian SSD formulae are obtained according to various criteria.
In Section \ref{sec:app}, we illustrate the application of the derived Bayesian SSD formulae in rare-disease trials for cases of known and unknown variances, respectively. This is followed by a performance evaluation of the proposed methodology in Section \ref{sec:sims}.
We conclude in Section \ref{sec:discuss} with a discussion about potential extensions of our methodology for a pragmatic while flexible application in modern clinical trials.

\section{Methods}
\label{sec:methods}

Suppose in the new experiment 
the difference $\mu_\Delta = \mu_{A} - \mu_{B} > 0$ indicates that $A$ is superior than $B$. In the context of clinical trials, for example, $A$ and $B$ can be treatments or interventions.
Let $X_{ij}$ be the observation from experimental unit $i$ assigned to group $j = A, B$. 
These observations are assumed to be i.i.d. normally distributed with mean $\mu_{j}$ and common variance $\sigma_{0}^2$. That is, for $i =1, \dots, n_j, \, j = A, B$, 
$$
X_{iA} \sim N(\mu_{A}, \sigma_{0}^2)  \quad \text{and}  \quad  X_{iB} \sim N(\mu_{B}, \sigma_{0}^2).
$$
We further suppose that a total of $K$ relevant historical datasets, denoted by $\boldsymbol{y}_1, \dots, \boldsymbol{y}_K$; each having the same data structure as that would accrue in the new experiment. Let $\theta_k$ denote the counterpart of $\mu_\Delta$, i.e., the difference in the normal means specific to each historical experiment, $k= 1, \dots, K$. 
In Section \ref{sec:borrowinfo}, we explain how these $\theta_1, \dots, \theta_K$, specific to the historical experiments, can be linked with the parameter $\mu_\Delta = \mu_A - \mu_B$ of inferential interest in the new experiment. With a brief review of commonly used criteria in Section \ref{sec:criteriaBayes}, we develop the Bayesian SSD approach in Section \ref{sec:SSDvar} for the new experiment, when leveraging pre-experimental information is an option.

\subsection{Borrowing of historical information from multiple sources}
\label{sec:borrowinfo}

\citet{BIOS:ZhengWason2019} propose a robust Bayesian model using commensurate priors \citep{BIOM:Hobbs2011, BA:Hobbs2012} to leverage complementary data from multiple sources for analysing a complex type of modern clinical trials. Specifically, data of complementary (sub)trials would be generated concurrently with that of a contemporary (sub)trial. 
Predictive priors for the contemporary parameter are specified to represent the complementary data, with a commensurate parameter introduced to determine the degree of borrowing from a specific source, i.e., a specific complementary (sub)study.

Following \citet{BIOS:ZhengWason2019}, we stipulate the commensurability, denoted by $\nu_k$, as the precision (i.e., inverse of variance) of a normal predictive distribution which is centred at the pre-experimental parameter $\theta_k, \, k = 1, \dots, K$. This leads to $K$ commensurate predictive distributions in the form of

\begin{equation}
\tilde{\theta}_k \mid \theta_k, \nu_k \sim N(\theta_k, \nu_k^{-1}),
\label{eq:commenprio}
\end{equation}

\noindent where each $\tilde{\theta}_k$ is regarded as {\em equivalent} to $\mu_\Delta$ in terms of the parameter space. More precisely, it means that the parameter space for $\tilde{\theta}_k$, as projected from a pre-experimental parameter $\theta_k$, would be defined with the same or comparable set of parameter values to that of $\mu_\Delta$. A spike-and-slab prior, which is a discrete mixture distribution, is placed on each $\nu_k$ for robust borrowing of information in the original proposal.  For analytic tractability in exact Bayesian inference, we consider using conjugate priors instead, i.e., a two-component Gamma mixture prior, for the predictive precision:
\begin{equation}
\nu_{k} \sim w_{k}\text{Gamma}(a_{01}, b_{01}) + (1-w_{k}) \text{Gamma}(a_{02}, b_{02}),
\label{eq:mixGamma}
\end{equation}
\noindent where $w_{k}$ denotes the prior mixture weight, on the scale of [0, 1], to represent preliminary scepticism about how commensurate $\theta_k$ and $\mu_\Delta$ would be.
The hyperparameters are chosen so that the first mixture component with $a_{01}, b_{01}$ has the density concentrated on small values of $\nu_k$, while the second mixture component with $a_{02}, b_{02}$ has density covering larger values of $\nu_k$. A large prior mixture weight allocated to either component distribution would thus result in sufficient down-weighting (with no borrowing at all as one extreme) or strong borrowing of historical information (with fully pooling as the other extreme), respectively. Stipulating with $0<w_k<1$ in \eqref{eq:mixGamma} produces a compromise between the two extreme cases. The strength of this Gamma mixture prior is then tuned by $w_k$, which can be interpreted as the prior probability of incommensurability. 

The joint pdf of $\tilde{\theta}_k$ and $\nu_{k}$, given information on $\theta_k$ that reveals the average difference between group $A$ relative to $B$ in a historical experiment $k$, will then be:
\begin{equation}
\begin{split}
f(\tilde{\theta}_k, \nu_{k} \mid \theta_k) &= f(\tilde{\theta}_k \mid \theta_k, \nu_{k}) g(\nu_{k})  \\ 
& = w_{k}\frac{b_{01}^{a_{01}}}{\sqrt{2\pi}\Gamma(a_{01})}\nu_{k}^{a_{01} - \frac{1}{2}} \exp\left( - \frac{2b_{01}\nu_{k}+\nu_{k}(\tilde{\theta}_k - \theta_k)^2}{2}\right) + \\
& \qquad  (1 - w_{k})\frac{b_{02}^{a_{02}}}{\sqrt{2\pi}\Gamma(a_{02})}\nu_{k}^{a_{02} - \frac{1}{2}} \exp\left( - \frac{2b_{02}\nu_{k} + \nu_{k}(\tilde{\theta}_k - \theta_k)^2}{2}\right),
\end{split}
\end{equation}

\noindent pointing to a Normal-Gamma mixture distribution.
We marginalise this mixture distribution for $\tilde{\theta}_k$ by integrating out the nuisance parameter $\nu_{k}$, and obtain
\begin{equation}
\begin{split}
f(\tilde{\theta}_k \mid \theta_k) & = \int f(\tilde{\theta}_k, \nu_{k} \mid \theta_k) {\rm d} \nu_{k} \\
& \propto w_{k} \left( \frac{(\tilde{\theta}_k - \theta_k)^2}{2b_{01}} + 1 \right)^{-\frac{2a_{01}+1}{2}} +  (1 - w_{k}) \left( \frac{(\tilde{\theta}_k - \theta_k)^2}{2b_{02}} + 1 \right)^{-\frac{2a_{02}+1}{2}},
\end{split}
\label{eq:tmix}
\end{equation}

\noindent which is a two-component mixture of non-standardised (shifted and scaled) $t$ distributions. In particular, the component $t$ distributions have their location parameters identically as $\theta_k$ yet scale parameters as $\frac{b_{01}}{a_{01}}$ and $\frac{b_{02}}{a_{02}}$, respectively.
Detailed derivation of \eqref{eq:tmix} and the demonstration of it being a non-standardised $t$ mixture distribution are given in Section A of the Web-based Supplementary Materials.
For easing the synthesis of $K$ predictive priors later on, we approximate this unimodal $t$ mixture distribution by a normal distribution that
\begin{equation}
\tilde{\theta}_k \mid \theta_k \, \dot\sim \, N\left(\theta_k,\frac{w_{k}b_{01}}{a_{01}-1} + \frac{(1-w_{k})b_{02}}{a_{02}-1} \right), \quad \text{ with } a_{01}, a_{02} > 1.
\label{eq:margtk}
\end{equation}

\noindent This approximation is based on the first two moments of the non-standardised $t$ mixture distribution, which are analytically available; see Section B of the Supplementary Materials for details.
The variance of the normal approximation, takes account of the dispersion of both $t$ mixture components.
The goodness of such normal approximation to the original $t$ mixture distribution depends on the degrees of freedom, $2a_{01}$ and $2a_{02}$, and the scale parameters, $\frac{b_{01}}{a_{01}}$ and $\frac{b_{02}}{a_{02}}$, which are of the investigators' choice. 
We show the numerical accuracy of this approximation in Section C of the Supplementary Materials.
Meanwhile, we note that the motivation is rather practical: it can bring considerable convenience for deriving a normal collective prior for $\mu_\Delta$, which represents information on $\theta_1, \dots, \theta_K$, by the convolution operator \citep{Grinstead1997}. 

With the normal approximation given by \eqref{eq:margtk}, we stipulate $\mu_\Delta$ as a linear combination of $K\geq 2$ hypothetical random variables, $\tilde{\theta}_k$, projected from the pre-experimental parameters. That is, $\mu_\Delta = \sum_k p_k \tilde{\theta}_k$, for $k = 1, \dots, K$. In particular,  the weights $p_1, \dots, p_K$ sum to 1, with each reflecting the relative importance of a corresponding pre-experimental dataset to constitute the collective predictive prior for $\mu_\Delta$. 
We note these weights could be associated with the prior mixture weights $1-w_k$, which describe our prior belief in the commensurability between a pre-experimental dataset and that to be collected from the new experiment. Applying the convolution operator for the sum of normal random variables, $\mu_\Delta$ has a normal prior distribution. Suppose that for $k=1, \dots, K$, each pre-experimental dataset leads to an estimate of $\theta_k \mid \boldsymbol{y}_k \sim N(m_k, s_k^2)$. The normal collective prior for $\mu_\Delta$ thus has the form of
\begin{equation}
\mu_\Delta \mid \boldsymbol{y}_1, \dots, \boldsymbol{y}_K \sim N\left(\sum_k p_k \lambda_k, \sum_k p_k^2\xi_k^2\right),
\label{eq:combtq}
\end{equation}

\noindent with
$$
\lambda_k = m_k \quad \text{ and } \quad \xi_k^2 = s_k^2 + \frac{w_{k}b_{01}}{a_{01}-1} + \frac{(1-w_{k})b_{02}}{a_{02}-1}, \quad (a_{01}, a_{02}>1)
$$
being the marginal prior means and variances.
It accounts for both the variability in a pre-experimental dataset $\boldsymbol{y}_k$ and the postulated level of incommensurability, $w_k$, through the Gamma mixture prior placed on the predictive precision, $\nu_k$.
We give more details in Section D of the Supplementary Materials for this derivation.
Using Bayes' Theorem, this collective prior will be updated by the new experimental data, denoted by $\boldsymbol{y}_{K+1}$, to a robust posterior $f_p(\mu_\Delta \mid \boldsymbol{y}_1, \dots, \boldsymbol{y}_K, \boldsymbol{y}_{K+1})$. Robustness is achieved through the use of a mixture of an informative and diffuse priors in the construction of the model. 

\subsection{Criteria for the Bayesian sample size determination}
\label{sec:criteriaBayes}

Most Bayesian SSD criteria aim to control a certain property of the posterior $f_p(\mu_\Delta \mid \boldsymbol{y}_1, \dots, \boldsymbol{y}_K, \boldsymbol{y}_{K+1})$, wherein the new experimental data $\boldsymbol{y}_{K+1}$ are unobserved at the design stage.  It is important to state at the outset that uncertainty of sampling a set of data, $\boldsymbol{y}_{K+1}$, from the entire {\em probability space} needs to be accounted for. In other words, there is no unique outcome or result of the new experiment. Thus, strictly, the Bayesian SSD criteria can only maintain the average properties of the posterior.
In what follows, we review some widely-used Bayesian SSD criteria for deriving our formulae accordingly.

\citet{JRSSD:Joseph1997} propose specifying a density region, $R(\boldsymbol{y}_{K+1})$, set to be bounded by $r$ and $r+\ell_0$ to contain possible parameter values. Here, $\ell_0$ is the desired interval length and $r$ chosen so that $R(\boldsymbol{y}_{K+1})$ is the highest posterior density (HPD) interval; so-called HPD because this interval includes the mode of the posterior distribution.
In particular, their specification is to ensure the coverage probability of $R(\boldsymbol{y}_{K+1})$ to be at least $1-\alpha$, when averaged over all possible samples. Formally, it requires that 
\begin{equation}
\int_\mathcal{Y}\left\{ \int_r^{r+\ell_0} f_p (\mu_\Delta \mid \boldsymbol{y}_1, \dots, \boldsymbol{y}_K, \boldsymbol{y}_{K+1}) {\rm d}\mu_\Delta \right\} f_d(\boldsymbol{y}_{K+1}) {\rm d} \boldsymbol{y}_{K+1} \geq 1- \alpha,
\label{ACC}
\end{equation}
\noindent where $\mathcal{Y}$ denotes the probability space and $f_d(\boldsymbol{y}_{K+1})$ the marginal distribution of the sample, i.e., the new experimental data. This Bayesian SSD criterion is generally applicable to both symmetric and asymmetric posterior distributions.
For the property of controlling the coverage probability, it is often referred to as the average coverage criterion (ACC). 
The posterior distribution in our context would be unimodal and symmetric about the posterior mean, as we can envisage from the collective prior given by \eqref{eq:combtq} which leverages pre-experimental information. We would then simply stipulate the HPD interval as
$$
R(\boldsymbol{y}_{K+1}) = \mathbb{E}(\mu_\Delta \mid \boldsymbol{y}_1, \dots, \boldsymbol{y}_K, \boldsymbol{y}_{K+1}) \pm \frac{\ell_0}{2},
$$
which coincides with the alpha-expectation tolerance region by \citet{AMS:Fraser1956}.

As an alternative to the ACC, one may want to fix the coverage of a posterior interval as $(1-\alpha_0)$ for the SSD, while limiting the interval length to be at most $\ell$. This proposal is commonly known as the average length criterion (ALC), with its first formal description given by \citet{JRSSD:Joseph1997}. Let $\ell'(\boldsymbol{y}_{K+1})$ be the random interval length of the posterior credible interval dependent on the unobserved new experimental data. Targeting a fixed coverage probability at level $(1-\alpha_0)$, one may solve $\ell'(\boldsymbol{y}_{K+1})$ to meet
$$
\int_r^{r+\ell'(\boldsymbol{y}_{K+1})} f_p (\mu_\Delta \mid  \boldsymbol{y}_1, \dots, \boldsymbol{y}_K, \boldsymbol{y}_{K+1}) {\rm d} \mu_\Delta = 1 - \alpha_0,
$$
where $r$ would be specified to give the HPD interval as that for the ACC above.
Averaged over all possible samples, the ALC requires that
\begin{equation}
\int_\mathcal{Y} \ell'(\boldsymbol{y}_{K+1}) f_d(\boldsymbol{y}_{K+1}) {\rm d} \boldsymbol{y}_{K+1} \leq \ell.
\label{ALC}
\end{equation} 
\noindent The ALC could be more favoured than the ACC, since most Bayesian practitioners are keen to report the average interval length of a fixed coverage probability, say, 95\%, in their data analyses.

As we can see, sample sizes chosen to meet the ACC by \eqref{ACC} or ALC by \eqref{ALC} rely on the marginal distribution of the new experimental data $\boldsymbol{y}_{K+1}$; that is, 
$$
f_d(\boldsymbol{y}_{K+1}) = \int f(\boldsymbol{y}_{K+1} \mid \mu_\Delta)\pi (\mu_\Delta) {\rm d} \mu_\Delta.
$$ 
When this predictive distribution of data also depends on nuisance parameters, say, the variance $\sigma_0^2$ being unknown, we may remove the dependence on the variance by integrating it out; then 
$$
f_d(\boldsymbol{y}_{K+1}) = \int \int f(\boldsymbol{y}_{K+1} \mid \mu_\Delta, \sigma_0^2)\pi (\mu_\Delta)g(\sigma_0^2){\rm d} \mu_\Delta {\rm d}\sigma_0^2.
$$ 
In our context, the prior distribution(s) for unknown parameter $\mu_\Delta$ as well as possibly unknown nuisance parameter $\sigma_0^2$ would be specified using pre-experimental information. The predictive distribution of data $\boldsymbol{y}_{K+1}$ would thus formally be $f_d(\boldsymbol{y}_{K+1} \mid \boldsymbol{y}_{1}, \dots, \boldsymbol{y}_{K})$, given our $\pi (\mu_\Delta \mid \boldsymbol{y}_{1}, \dots, \boldsymbol{y}_{K})$ and $g(\sigma_0^2 \mid \boldsymbol{y}_{1}, \dots, \boldsymbol{y}_{K})$.

Both the ACC and ALC are based on defining a probability measure of the posterior distribution.
We consider one additional criterion that relates to the moments of the posterior distribution. For practicality reasons, we focus on the precision of the posterior mean estimate $\mu_\Delta$, i.e., the second central moment of the posterior probability distribution. This would be referred to as the average posterior variance criterion (APVC) hereafter. Given a fixed level of dispersion $\epsilon_0$, a suitable sample size is chosen to ensure that
$$
\mathbb{E}_\mathcal{Y}[\text{Var}(\mu_\Delta \mid \boldsymbol{y}_1, \dots, \boldsymbol{y}_K, \boldsymbol{y}_{K+1})] \leq \epsilon_0.
$$
As \citet{JRSSD:Adcock1997} commented, this criterion is equivalent to using the $L_2$-norm loss function for inferences: $L_2 (\mu_\Delta) = (\mu_\Delta - \mathbb{E}(\mu_\Delta \mid \boldsymbol{y}_1, \dots, \boldsymbol{y}_K, \boldsymbol{y}_{K+1}))^2$.
It is also worth noting that the literature has documented many other Bayesian approaches to SSD, e.g., based on the use of utility theory \citep{JRSSD:Lindley1997} and Bayes factors \citep{JRSSD:Weiss1997}. The reader may be interested in working out more Bayesian SSD formulae based on our posterior distribution while applying alternative criteria.

To complete the brief review of Bayesian SSD criteria of our selection, we note that the fixed values of $\ell_0$, $\alpha_0$ and $\epsilon_0$ are all positive real numbers. How these values would affect the sample size chosen for planning a new experiment will be investigated through numerical evaluation in Section \ref{sec:sims}.

\subsection{Sample size required for comparing two normal means}
\label{sec:SSDvar}

Denote the groupwise sample sizes in the new experiment by $n_A$ and $n_B$, respectively.
For most settings, the experimental data $\boldsymbol{y}_{K+1}$ contain two independent random vectors $(X_{1A}, \dots, X_{n_A A})$ and $(X_{1B}, \dots, X_{n_B B})$, sampled from the populations that are assumed to have a common variance, denoted by $\sigma_0^2$. 

\subsubsection*{For cases of known variance} 

If the common variance $\sigma_0^2$ is known exactly, the sample means $\bar{x}_{A} \mid \mu_A \sim N(\mu_A, \frac{\sigma_0^2}{n_A})$ and $\bar{x}_B \mid \mu_B \sim N(\mu_B, \frac{\sigma_0^2}{n_B})$. This  further leads to 
$\bar{x}_\Delta \mid \mu_A, \mu_B \sim N\left(\mu_\Delta, \frac{\sigma_0^2}{n_A} + \frac{\sigma_0^2}{n_B}\right)$,
where $\bar{x}_\Delta = \bar{x}_A - \bar{x}_B$, and $\mu_\Delta  = \mu_A - \mu_B$ that has a normal prior based on pre-experimental datasets $\boldsymbol{y}_1, \dots, \boldsymbol{y}_K$, as was given in \eqref{eq:combtq}. 
Since the joint likelihood of the $n_A + n_B$ measurements in the new experiment $\mathcal{L}(\boldsymbol{y}_{K+1} | \mu_A, \mu_B) \propto \mathcal{L}(\bar{x}_\Delta | \mu_A, \mu_B)$, we formulate the data likelihood in terms of $\bar{x}_\Delta$ which is regarded as a random variable. We further derive the posterior distribution as
\begin{equation}
\mu_\Delta \mid \boldsymbol{y}_1, \dots, \boldsymbol{y}_K, \boldsymbol{y}_{K+1} \sim N \left(\eta, \left( \frac{1}{\sum p_k^2 \xi_k^2} + \frac{1}{\left(\frac{1}{n_A} + \frac{1}{n_B}\right)\sigma_0^2} \right)^{-1} \right),
\label{eq:postknvar}
\end{equation}

\noindent where 
$$
\eta = \frac{\left(\frac{1}{n_A} + \frac{1}{n_B}\right)\sigma_0^2}{\sum p_k^2 \xi_k^2 + \left(\frac{1}{n_A} + \frac{1}{n_B}\right)\sigma_0^2} \sum p_k \lambda_k + \frac{\sum p_k^2 \xi_k^2}{\sum p_k^2 \xi_k^2 + \left(\frac{1}{n_A} + \frac{1}{n_B}\right)\sigma_0^2}\bar{x}_{\Delta}.
$$
The marginal distribution (unconditional on $\mu_\Delta$) for the difference in sample means is given by
$$
\bar{x}_\Delta \mid \boldsymbol{y}_1, \dots, \boldsymbol{y}_K \sim N\left(\sum_k p_k \lambda_k, \left(\frac{1}{n_A} + \frac{1}{n_B}\right)\sigma_0^2 + \sum_k p_k^2\xi_k^2\right),
$$
which corresponds to the marginal distribution of the new experimental data, $f_d(\boldsymbol{y}_{K+1}\mid \boldsymbol{y}_{1}, \dots, \boldsymbol{y}_{K})$, in Section \ref{sec:criteriaBayes}.

As \citet{JRSSD:Joseph1997} noted, the ACC and ALC result in the same outcome for cases where the variance is known. Hence, we illustrate using the ACC in the following. Letting the HPD interval $(r, r+\ell_0)$ stretch symmetrically around the posterior mean $\eta$, the coverage can be computed by
$$
\mathbb{P}\left[|\mu_\Delta - \eta| \leq \frac{\ell_0}{2} \mid \boldsymbol{y}_1, \dots, \boldsymbol{y}_K, \boldsymbol{y}_{K+1}  \right] = \Phi \left( \sqrt{\frac{1}{\sum p_k^2 \xi_k^2} + \frac{1}{\left(\frac{1}{n_A} + \frac{1}{n_B}\right)\sigma_0^2}} \frac{\ell_0}{2} \right),
$$
where $\Phi(\cdot)$ is the cumulative distribution function of the standard normal distribution.
We thus have 
$$
\Phi \left( \sqrt{ \frac{1}{\sum p_k^2 \xi_k^2} + \frac{1}{\left(\frac{1}{n_A} + \frac{1}{n_B}\right)\sigma_0^2}} \frac{\ell_0}{2} \right) \geq 1 - \alpha,
$$
leading to 
\begin{equation}
\frac{n_A n_B}{n_A + n_B} \geq \left( \frac{4 z^2_{\alpha/2}}{\ell_0^2} - \frac{1}{\sum p_k^2 \xi_k^2} \right)\sigma_0^2,
\label{eq:ACCkvar}
\end{equation}

\noindent where $z_{\alpha/2}$ is the upper $(\alpha/2)$-th quantile of the standard normal distribution, i.e., $\Phi^{-1}(1- \alpha/2)$. 
Similarly, averaging over the entire data space, the APVC gives
\begin{equation}
\frac{n_A n_B}{n_A + n_B} \geq \left( \frac{1}{\epsilon_0} -  \frac{1}{\sum p_k^2 \xi_k^2} \right)\sigma_0^2.
\end{equation}

\noindent In theory, when $\sum p_k^2\xi_k^2$, the prior variance for $\mu_\Delta \mid \boldsymbol{y}_1, \dots, \boldsymbol{y}_K$, is so small that the right-hand side of the inequalities above becomes zero or negative, there is no need to conduct a new experiment for collecting more information. 

\subsubsection*{For cases of unknown variance}

When the common variance $\sigma_0^2$ is unknown, we assume that the quantity $c\sum p_k^2 \xi_k^2/\sigma_0^2\sim \chi^2{(c)}$, where $\chi^2{(c)}$ refers to a Chi-square distribution with $c$ degrees of freedom \citep{Gelman2013}. 
This is equivalent to the prior specification that $\sigma_0^2 \sim $ Inv-Gamma($\frac{c}{2}, \frac{c\sum p_k^2\xi_k^2}{2}$); hence, the larger value $c$ takes, the more the unknown $\sigma_0^2$ converges to the prior variance for $\mu_\Delta \mid \boldsymbol{y}_1, \dots, \boldsymbol{y}_K$.
The marginal posterior for $\mu_\Delta$ will then be obtained by intergrating out the nuisance parameter $\sigma_0^2$:
\begin{equation}
\begin{split}
f_p (\mu_\Delta \mid \boldsymbol{y}_1, \dots, \boldsymbol{y}_K, \boldsymbol{y}_{K+1}) & = \int \pi_p (\mu_\Delta, \sigma_0^2 \mid  \boldsymbol{y}_1, \dots, \boldsymbol{y}_K, \boldsymbol{y}_{K+1}) g(\sigma_0^2) {\rm d} \sigma_0^2 \\
& \propto \exp \left(-\frac{(\mu_\Delta - \sum p_k \lambda_k)^2}{2 \sum p_k^2 \xi_k^2} \right) \left[ 1 + \frac{1}{c}\cdot \frac{ (\mu_\Delta - \bar{x}_\Delta)^2}{\left( \frac{1}{n_A} + \frac{1}{n_B} \right) \sum p_k^2 \xi_k^2}\right]^{-\frac{c+1}{2}},
\end{split}
\label{mupostI}
\end{equation}

\noindent that is, the posterior is proportional to the product of normal and non-standardised $t$ kernels \citep{Ahsanullah2014}.
Detailed steps for deriving \eqref{mupostI} are given in Section E of the Supplementary Materials. 
In particular, the $t$ density kernel (with the location and scale parameters being $\bar{x}_\Delta$ and $\left(\frac{1}{n_A} + \frac{1}{n_B} \right)\sum p_k^2 \xi_k^2$, respectively) can be related to a normal kernel with the same location parameter and the variance as $\left(\frac{1}{n_A} + \frac{1}{n_B} \right)\sigma_0^2$, conditional on $c\sum p_k^2 \xi_k^2/\sigma_0^2 \sim \chi^2 (c)$.
The posterior \eqref{mupostI} can thus be further developed as
\begin{equation}
\begin{split}
f_p (\mu_\Delta \mid \sigma_0^2, \boldsymbol{y}_1, \dots, \boldsymbol{y}_K, \boldsymbol{y}_{K+1}) & \propto \exp \left(-\frac{(\mu_\Delta - \sum p_k \lambda_k)^2}{2 \sum p_k^2 \xi_k^2} \right) \exp \left( - \frac{(\mu_\Delta - \bar{x}_\Delta)^2}{2\left( \frac{1}{n_A} + \frac{1}{n_B} \right)\sigma_0^2} \right) \\
& \overset{\mathrm{def}}{=} \exp \left( - \frac{(\mu_\Delta - \mu_N)^2}{2\sigma_N^2} \right),
\end{split}
\label{mupostII}
\end{equation}
\noindent with 
\begin{equation*}
\mu_N = \frac{\left(\frac{1}{n_A} + \frac{1}{n_B}\right)\sigma_0^2}{\sum p_k^2 \xi_k^2 + \left(\frac{1}{n_A} + \frac{1}{n_B}\right)\sigma_0^2} \sum p_k \lambda_k + \frac{\sum p_k^2 \xi_k^2}{\sum p_k^2 \xi_k^2 + \left(\frac{1}{n_A} + \frac{1}{n_B}\right)\sigma_0^2}\bar{x}_{\Delta}
\end{equation*}
\noindent and 
\begin{equation*}
\sigma_N^2 = \left( \frac{1}{\sum p_k^2 \xi_k^2} + \frac{1}{\left(\frac{1}{n_A} + \frac{1}{n_B}\right)\sigma_0^2} \right)^{-1},
\end{equation*}
\noindent which is consistent with \eqref{eq:postknvar} but here with unknown $\sigma_0^2 \sim$ Inv-Gamma($\frac{c}{2}, \frac{c\sum p_k^2\xi_k^2}{2}$). 
We can also find the distribution for $\bar{x}_\Delta$ unconditional on $\mu_\Delta$ as
$$
\bar{x}_\Delta \mid \sigma_0^2, \boldsymbol{y}_1, \dots, \boldsymbol{y}_K \sim N\left( \sum p_k \lambda_k, \left(\frac{1}{n_A} + \frac{1}{n_B}\right)\sigma_0^2 + \sum_k p_k^2\xi_k^2 \right);
$$
see Section E of the Supplementary Materials for the derivation.
Apparently, this marginal distribution for $\bar{x}_\Delta$ relies on prior distribution for the unknown $\sigma_0^2$, which may yield different solutions of $n_A$ and $n_B$ across the Bayesian SSD criteria considered in this paper.

Let the interval $(a, a+\ell_0)$ be symmetric about $\mu_N$ given the marginal posterior for $\mu_\Delta$ in \eqref{mupostII}. When applying the ACC, the sample size is found requiring 

\begin{equation*}
\mathbb{P}\left[ |\mu_\Delta - \mu_N| \leq \frac{\ell_0}{2} \mid \boldsymbol{y}_1, \dots, \boldsymbol{y}_K, \boldsymbol{y}_{K+1} \right] = \mathbb{P}\left[ \frac{1}{\sigma_N}|\mu_\Delta - \mu_N| \leq \frac{\ell_0}{2\sigma_N} \mid \boldsymbol{y}_1, \dots, \boldsymbol{y}_K, \boldsymbol{y}_{K+1} \right] \geq 1-\alpha;
\end{equation*}

\noindent thus

\begin{equation*}
\frac{\ell_0}{2\sigma_N} = \sqrt{\frac{1}{\sum p_k^2 \xi_k^2} + \frac{1}{\left(\frac{1}{n_A} + \frac{1}{n_B}\right)\sigma_0^2}}\cdot \frac{\ell_0}{2} \geq z_{\alpha/2},
\end{equation*}

\noindent where $z_{\alpha/2}$ denotes the upper $(\alpha/2)$-th quantile of the standard normal distribution. We rewrite the expression and obtain 

\begin{equation}
\frac{n_A n_B}{n_A + n_B} \geq \left(\frac{4 z_{\alpha/2}^2}{\ell_0^2} - \frac{1}{\sum p_k^2 \xi_k^2} \right)\int_0^\infty \sigma_0^2 g(\sigma_0^2){\rm d}\sigma_0^2,
\label{eq:ACCunkvar}
\end{equation}

\noindent where $g(\sigma_0^2)$ is the pdf of an Inv-Gamma($\frac{c}{2}, \frac{c\sum p_k^2 \xi_k^2}{2}$) distribution. The reader may compare this inequality with what was obtained for cases where $\sigma_0^2$ is known in \eqref{eq:ACCkvar}.

For computing the ALC sample size, we would have to average the random credible interval length $\ell'(\bar{x}_\Delta) = 2z_{\alpha_0/2}\sigma_N$ over the marginal distribution for $\bar{x}_\Delta$ which varies with $\sigma_0^2$. According to the definition of ALC, we obtain that

\begin{equation}
2z_{\alpha_0/2} \int_0^\infty \left(\frac{1}{\sum p_k^2 \xi_k^2} + \frac{1}{\left(\frac{1}{n_A} + \frac{1}{n_B}\right)\sigma_0^2}\right)^{-\frac{1}{2}} g(\sigma_0^2){\rm d}\sigma_0^2 \leq \ell,
\label{eq:ALCunkvar}
\end{equation}

\noindent which does not have a closed-form solution. 
This require a search over the integers for $n_A$ and $n_B$ to find the smallest sum that satisfies the inequality.
With the use of APVC, we would likewise remove the dependence on $\sigma_0^2$ by intergration. The formula thus becomes
 
\begin{equation}
\frac{n_A n_B}{n_A + n_B}\geq \left(\frac{1}{\epsilon_0} - \frac{1}{\sum p_k^2 \xi_k^2} \right)\int_0^\infty \sigma_0^2 g(\sigma_0^2){\rm d}\sigma_0^2.
\label{eq:APVCunkvar}
\end{equation}

\noindent Finally, we note that for computing an exact sample size for the new experiment, the allocation ratio $n_A:n_B$ would have to be specified, for example, considering an equal allocation by setting $n_A:n_B = 1:1$.

\section{Application}
\label{sec:app}

In this section, we illustrate the application of our Bayesian SSD formulae to rare-disease trials, for which it is generally infeasible to enroll a large sample size based on frequentist SSD formulae. 
Such rare-disease trials are unlikely to be conducted without any preceding investigation. Pre-trial information collected from relevant studies (for example, historical clinical trials evaluating the same treatment in a similar patient population) or elicited from expert opinion, would often be available. Our proposed Bayesian SSD methodology provides a framework to formally utilise this prior information for planning a new trial.

\citet{SiM:Hampson2014} present a Bayesian approach for elicitation of expert opinion on the parameters for enhanced design and analysis of rare-disease trials. A two-day elicitation meeting \citep{PONE:Hampson2015} was held for the MYPAN trial, which compares the efficacy of a new treatment (labelled $A$) relative to the standard of care (labelled $B$) for polyarteritis nodosa, a rare and severe inflammatory blood vessel disease. 
Priors were elicited from the input of 15 experts individually.
Specifically, opinion was sought on (i) the probability that a patient given $B$ would achieve disease remission within 6 months (a dichotomous event), and (ii) the log-odds ratio of remission rates. Consensus distributions for the remission rates were obtained, with the mode at 71\% for $A$ and 74\% for $B$. In what follows, we regard expert opinion as a type of pre-trial information, and generate hypothetical examples considering characteristics of the MYPAN trial to illustrate the application of the proposed Bayesian SSD approach.

In line with the original assumptions of the MYPAN trial, we suppose the log-odds ratio of treatment benefit, $\theta_k = \log(\rho_{Ak}(1- \rho_{Bk}))/((1-\rho_{Ak})\rho_{Bk})$, can be adequately modelled by a normal distribution. Here, $\rho_{jk}$ denotes the probability of remission for a patient receiving treatment $j = A, B$, respectively. 
Furthermore, we assume some expert opinion had been expressed, as summarised in the form of $\theta_k\mid \boldsymbol{y}_k \sim N(m_k, s_k^2), \, k = 1, \dots, K$. Eliciting such expert opinion is a non-trivial problem; we refer the reader to the statistical literature on elicitation of prior distributions in Bayesian inferences \citep{Elicitation2017}.
For illustration, we assume there were five sets of expert opinion that had been summarised as $N(-0.26, 0.25), \, N(-0.24, 0.23), \, N(-0.37, 0.22)$, $N(-0.34, 0.36)$ and $N(-0.32, 0.26)$.
Opinion would also be sought on the experts' skepticism about the predictability of each pre-trial parameter $\theta_k$ towards the parameter $\mu_\Delta$, measured on the continuous scale of 0 to 1, to specify the prior probabilities of incommensurability, $w_k, \, k = 1, \dots, 5$. 
In this example, we suppose such pre-trial information may be valued about equally, so stipulate $w_1 = 0.15,\, w_2 = 0.20, \, w_3 = 0.17, \, w_4 = 0.13, \, w_5 = 0.20$ for robust borrowing of information. 
Pragmatically, the trial statistician could look into the levels of pairwise commensurability between the $N(m_k, s_k^2)$ distributions based on distributional discrepancy, such as Hellinger distance \citep{Hellinger:Dey1994}, to reconcile the choices of value for $w_k$.

For reaching a collective prior for $\mu_\Delta \mid \boldsymbol{y}_1, \dots, \boldsymbol{y}_5$, weights $p_1, \dots, p_5$ need to be specified to reflect the relative commensurability of each set of pre-trial information with data to be accrued in the new trial. Both $w_k$ and $p_k$ might be determined by some distance measure between parameters $\theta_k$ and $\mu_\Delta$ \citep{BIOS:ZhengWason2019}, for the rationale that we hope to discount pre-experimental information to a larger extent when it would be less similar with the new experimental data. For implementing the proposed Bayesian SSD approach in this paper, we compute the weights as 
$$
p_k = \frac{\exp(-w_k^2/s_0)}{\sum_k \exp(-w_k^2/s_0)},
$$
where $s_0$ in this descreasing function of $w_k$ determines how concentrated the weights $p_1, \dots, p_K$ would be around the average, $1/K$. With a value of $s_0 \gg w_k$, nearly all $p_k$ converge to $1/K$ irrespective of the values of $w_k$. Whereas, with $s_0 \rightarrow 0^+$, the smallest $w_k$ would have $p_k \rightarrow 1$, meaning that the corresponding $\theta_k \mid \boldsymbol{y}_k$ tends to dominate the collective prior. For illustration, we let $s_0 = 0.05$. Thus, with the $w_k$ specified above, we obtain $p_1 = 0.23, \, p_2 = 0.16, \, p_3 = 0.20, \, p_4 = 0.25, \, p_5 = 0.16$.

We let the prior predictive precision $\nu_k \sim w_k \text{Gamma}(2, 2) + (1-w_k) \text{Gamma}(18, 3)$, where the means together with 95\% credible interval of the component Gamma distributions are 1.000 (0.121, 2.786) and 6.000 (3.556, 9.073), respectively. 
This mixture prior for $\nu_k$ is specified to fully account for two extreme cases as either no borrowing or strong borrowing of pre-experimental information, which correspond to setting $w_k = 1$ and $w_k = 0$, respectively.
In our modelling, this then leads to a collective prior $\mu_\Delta \mid \boldsymbol{y}_1, \dots, \boldsymbol{y}_5 \sim N(-0.309, 0.154)$, of which the prior 95\% credible interval symmetric about the prior mean is (-1.078, 0.460).
We assume known variance of $\sigma_0^2 = 0.35$ in the new trial under planning and suppose $n_A = n_B$. The total sample sizes (i.e., $n_A + n_B$) found based on the AAC and ALC criteria are then both 41.8 for 95\% posterior coverage probability and the credible interval length as 0.65 on average. For cases of unknown $\sigma_0^2$, we let $\sigma_0^2 \sim$ Inv-Gamma($2.500, 0.385$) (i.e., setting $c=5$). The ACC and ALC sample sizes become 30.7 and 24 for attaining the same posterior behaviours, respectively. 
Targeting $\epsilon_0 = 0.03$, the APVC sample sizes are 32.2 and 27.6 for known and unknown $\sigma_0^2$ (where we set $c=5$), respectively. 
Here, we have reported the sample sizes by rounding the result of an exact solution to one decimal place and by the smallest integer if found based on a search procedure.

In this illustrative example, sample sizes derived for cases of known $\sigma_0^2 = 0.35$ appear larger than those yielded by the same criterion for unknown $\sigma_0^2$. This is because by setting $c=5$, we in fact additionally permit certain amount of borrowing to inform the variance in the new study; more specifically, $\sigma_0^2$ would be thought of as similar to some extent to the prior variance of $\mu_\Delta \mid \boldsymbol{y}_1, \dots, \boldsymbol{y}_5$, say, 0.154, which is smaller than the fixed $\sigma_0^2 = 0.35$ in our illustration. We will examine the sensitivity of our formulae to the choice of $c$ and provide comprehensive interpretation in Section \ref{sec:sims}.

\section{Performance evaluation}
\label{sec:sims}

In this section, we provide a performance evaluation of the proposed formulae, with sample sizes computed and visualised under various configurations of parameters. 

\subsection{Basic settings}

Motivated by the MYPAN trial, we generate four base scenarios of historical data, which are configured with different levels of pairwise (in)commensurability and informativeness.
Such pre-experimental information from $K$ sources is supposed to have been summarised as $\theta_k \mid \boldsymbol{y}_k \sim N(m_k, s_k^2), \, k = 1, \dots, K$. 
For each configuration of hypothetical historical data, two distinct sets of robust weights I and II are considered to implement the proposed approach for borrowing of information. 
These robust weights are chosen for illustrative purposes to (a) reflect high and low level of prior confidence in the historical data when they are consistent between themselves, or (b) designate certain source of historical data to be more influential.
Table \ref{tab:SimSc} lists four base scenarios of historical data along with their assigned robust weights.
We compute the squared Hellinger distances of any two $N(m_k, s_k^2)$ distributions to describe their pairwise (in)commensurability, which are visualised in Figure S2 of the Web-based Supplementary Materials.
Prior probabilities of incommensurability, $w_k$, could be chosen at similar levels as such distances for leveraging each $N(m_k, s_k^2)$ in the new experiment. The values of $w_k$ in Table \ref{tab:SimSc} for our numerical study are justified as no greater than 0.500, as the largest squared Hellinger distance visualised in Figure S2 is below 0.500.
The weights, $p_k$, for prioritising certain prior information are determined following our stipulation in Section \ref{sec:app}.

In Table \ref{tab:SimSc}, the first two configurations of historical data represent situations of consistent pre-experimental information from different sources.  Robust weights I and II for these configurations consistently down-weight all historical data to a small and larger extent, respectively.
Configurations 3 and 4 represent situations of divergent historical data; each has their own robust weights I and II to downplay some informative historical data to a small and larger extent, respectively. 
The collective priors $N(\sum p_k \lambda_k, \sum p_k^2 \xi_k^2)$ are derived by specifying $\nu_k \sim w_k \text{Gamma}(2, 2) + (1-w_k) \text{Gamma}(18, 3)$. We note that the component Gamma distributions of the mixture prior can be essential, as choices are highly impactful on the prior effective sample size. Thus, in this numerical evaluation, we also examine how Bayesian SSD for the new trial would change given various Gamma mixture priors.

\begin{table}
\scriptsize
\centering
\caption{Configurations of hypothetical historical data, each accompanied by two sets of weights for robust borrowing of information. Pre-experimental information about $\theta_k \mid \boldsymbol{y}_k$ is assumed to have been summarised by a $N(m_k, s_k^2)$ prior for $k = 1, \dots, 5$.}
\begin{tabular}{@{\extracolsep{1.5pt}}llcccccccc@{}}
\toprule
& & &\multicolumn{5}{@{}c}{Hypothetical historical data}  & \multirow{2}{*}{$\sum p_k \lambda_k$}   & \multirow{2}{*}{$\sum p_k^2 \xi_k^2$}   \\
  \cline{4-8}  \\[-0.4em]
 &  & & $k=1$ & $k=2$  & $k=3$ & $k=4$  & $k=5$   \\
\midrule
Configuration 1 &  & $m_k$  & -0.260  & -0.240  & -0.370  & -0.340  & -0.320  \\
 &  & $s_k^2$    & 0.250  &  0.230  &  0.220  &  0.360  &  0.260  \\[0.2em]
   \cline{2-10}  \\[-0.4em]
& \multirow{2}{*}{Robust weights I}  & $w_k$  & 0.103  & 0.175  & 0.081  & 0.143  & 0.077  & \multirow{2}{*}{-0.311}  & \multirow{2}{*}{0.129}  \\
& & $p_k$ & 0.214  & 0.143  & 0.232  & 0.176  & 0.235  \\
 \cline{2-10} \\[-0.4em]
& \multirow{2}{*}{Robust weights II}  & $w_k$  & 0.252  & 0.319  & 0.140  & 0.306  & 0.149  & \multirow{2}{*}{-0.325}  & \multirow{2}{*}{0.198}  \\
& & $p_k$  & 0.149  & 0.069  & 0.359  & 0.082  & 0.341  \\
  \cline{1-10} \\[-0.4em]
Configuration 2 &  & $m_k$   & -0.260  & -0.240  & -0.370  & -0.340  & -0.320  \\
 &  & $s_k^2$ & 0.100  &  0.100  &  0.100  &  0.100  &  0.100    \\ [0.2em]
   \cline{2-10}  \\[-0.4em]
& \multirow{2}{*}{Robust weights I}  & $w_k$  & 0.103  & 0.175  & 0.081  & 0.143   & 0.077  & \multirow{2}{*}{-0.311}  & \multirow{2}{*}{0.096}  \\
& & $p_k$  & 0.214  & 0.143  & 0.232  & 0.176  & 0.235  \\
  \cline{2-10} \\[-0.4em]
 & \multirow{2}{*}{Robust weights II}  & $w_k$  & 0.252  & 0.319  & 0.140  & 0.306   & 0.149  & \multirow{2}{*}{-0.325}  & \multirow{2}{*}{0.158}  \\
 &  & $p_k$  & 0.149  & 0.069  & 0.359  & 0.082  & 0.341  \\
  \cline{1-10} \\[-0.4em]
Configuration 3 &  & $m_k$  & -0.260  & -0.170  & -0.440  & -0.150  & 0.120  \\
 &  & $s_k^2$     &  0.250  &  0.640  &  0.970  &  1.540  & 0.590   \\[0.2em]
   \cline{2-10}  \\[-0.4em]
& \multirow{2}{*}{Robust weights I}  & $w_k$  & 0.101  & 0.219  & 0.385  & 0.385   & 0.304  & \multirow{2}{*}{-0.198}  & \multirow{2}{*}{0.295} \\
& & $p_k$    & 0.559  & 0.263  & 0.035  & 0.035  & 0.108  \\
  \cline{2-10} \\[-0.4em]
 & \multirow{2}{*}{Robust weights II}  & $w_k$  & 0.325  & 0.203  & 0.171  & 0.180   & 0.272  & \multirow{2}{*}{-0.215}  & \multirow{2}{*}{0.379} \\
 &  & $p_k$   & 0.065  & 0.235  & 0.298  & 0.280  & 0.122  \\
  \cline{1-10} \\[-0.4em]
  Configuration 4 &  & $m_k$  & -0.260  & -0.170  & -0.440  & -0.150  & 0.120  \\
 &  & $s_k^2$    &  0.250  &  0.150  &  0.400  &  0.890  & 0.220   \\[0.2em]
   \cline{2-10}  \\[-0.4em]
& \multirow{2}{*}{Robust weights I}  & $w_k$  & 0.066  & 0.303  & 0.459  & 0.355   & 0.115  & \multirow{2}{*}{-0.099}  & \multirow{2}{*}{0.226} \\
& & $p_k$   & 0.473  & 0.082  & 0.008  & 0.041  & 0.396  \\
  \cline{2-10} \\[-0.4em]
 & \multirow{2}{*}{Robust weights II}  & $w_k$    & 0.537  & 0.306  & 0.054  & 0.220   & 0.350  & \multirow{2}{*}{-0.312}  & \multirow{2}{*}{0.343}  \\
 &  & $p_k$   & 0.002   & 0.098  & 0.602  & 0.243  & 0.055   \\
\bottomrule
\end{tabular}
\label{tab:SimSc}
\end{table}


We compare the sample sizes computed using the proposed Bayesian SSD formulae with those computed (a) without robustification, i.e., setting each $w_k = 0$ for $k = 1, \dots, 5$, (b) without leveraging historical information for $\mu_\Delta$, i.e., setting each $w_k = 1$, (c) from the proper Bayesian SSD approach driven by a single prior, here specified as the most informative $N(m_k, s_k^2)$, for example, $N(-0.37, 0.22)$ for configuration 1, and (d) from an optimal approach as the benchmark. Specifically, the optimal approach is coupled with a perfectly commensurate prior, by equating $\sigma_0^2$ to the collective prior variance $\sum p_k^2 \xi_k^2$. In this way, the corresponding result would serve as the benchmark referring to the scenario of perfect consistency between the collective prior and the new data, so the largest saving in sample size could be attained by using the proposed methodology. For cases of unknown $\sigma_0^2$, the optimal sample sizes could be approached by setting $c$ to a sufficiently large value. 

\subsection{Results}

Figure \ref{fig:SScomp1} visualises a subset of the results, which compare the proposed Bayesian SSD formulae using robust weights I and II with the alternative approaches for cases of known and unknown $\sigma_0^2$, respectively. 
Here, we assume $\sigma_0^2 = 0.35$ and, if unknown, $\sigma_0^2\sim$ Inv-Gamma(1.5, 1.5$\times \sum p_k^2 \xi_k^2$) for illustration. 
We fix the posterior credible interval length $\ell_0 = 0.65$ to find the ACC sample sizes, so that the average coverage probability would be 95\%, that is, targeting $\alpha = 0.05$ in \eqref{eq:ACCkvar}. Likewise, for computing the ALC sample sizes, we fix $\alpha_0 = 0.05$ and constrain the average length of the posterior credible interval below 0.65. When applying the APVC, sample sizes are found with the average posterior variance retained to level $\epsilon = 0.03$.


\begin{figure}[!ht]
\centering
\includegraphics[width=0.8\linewidth]{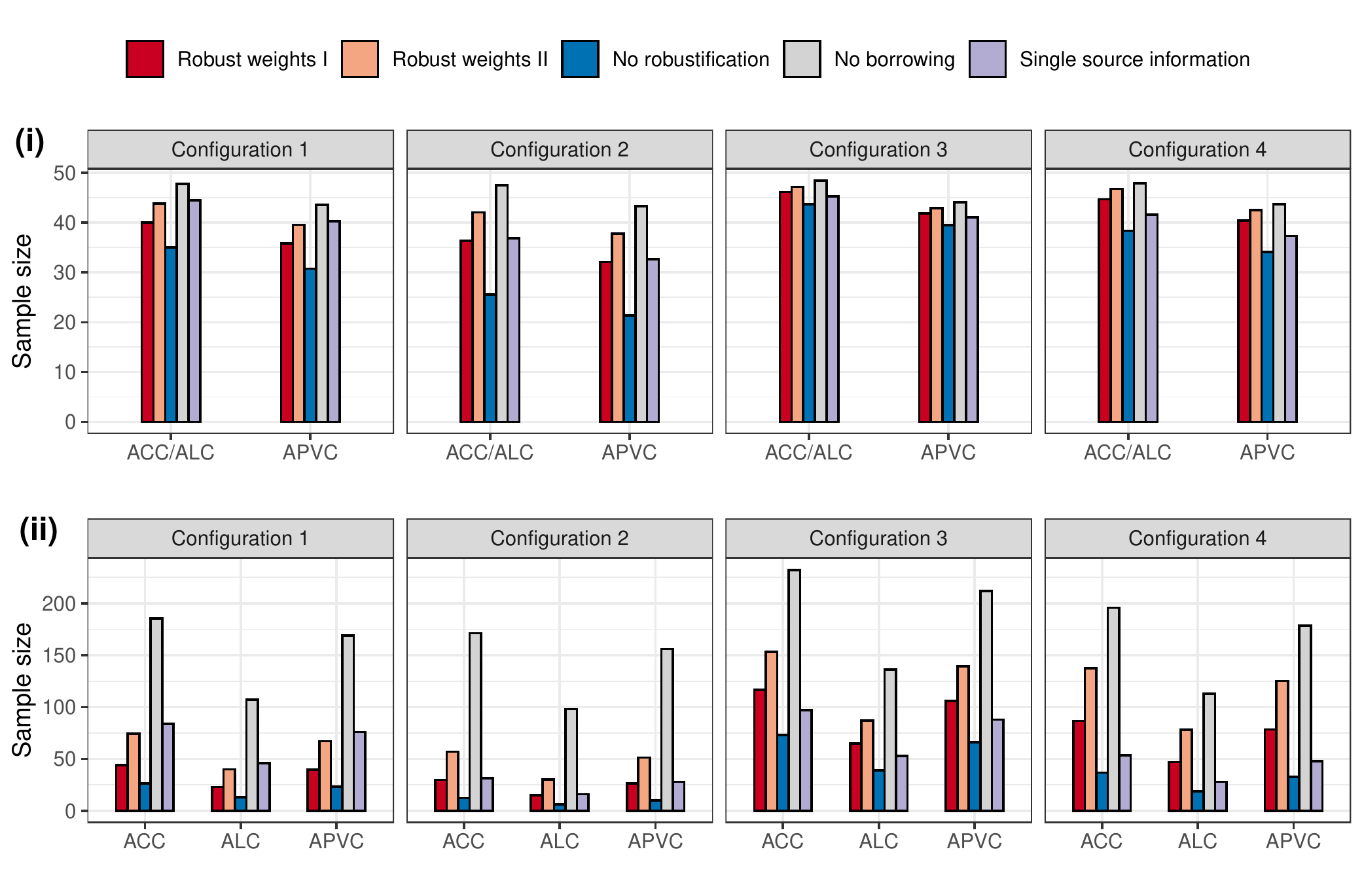}
\caption{Comparison of the Bayesian SSD approaches in terms of the sample size obtained according to the ACC, ALC and APVC criteria for cases of (i) known $\sigma_0^2 = 0.35$ and (ii) unknown $\sigma_0^2$. Sample sizes in subfigure (ii) for unknown $\sigma_0^2$ are computed setting $c=3$, i.e., assuming that $\sigma_0^2 \sim$ Inv-Gamma$(1.5, 1.5\times\sum p_k^2\xi_k^2)$, for fairly limited use of pre-experimental information to inform the variance $\sigma_0^2$.}
\label{fig:SScomp1}
\end{figure}

In all configurations 1 -- 4, we see that the sample sizes computed according to the same criterion, using robust weights I, are smaller than those using robust weights II. This is because following our setting the collective prior, produced by robust weights I, has smaller variance than its counterpart by robust weights II, for each configuration. 
Moreover, sample sizes yielded using either robust weights I or II are always bounded by those using no robustification ($w_k = 0$) and no borrowing ($w_k = 1$). We may think that no robustification leads to the least conservative result by the proposed SSD formulae, for the given historical information fully used.  These, however, are not necessarily identical to the optimal situations, where $\sigma_0^2$ is equated to the collective prior variance, or largely determined by the latter if unknown. In Figure \ref{fig:SScomp1}, we omit the benchmark optimal sample sizes that may be obtained by using the proposed formulae with robust weights I and II for each configuration. Yet we will comment on the maximal saving that the proposed SSD approach can achieve in the following along with other figures.

The height difference across bars of sample sizes, computed using our approach with robust weights I or II and no borrowing ($w_k = 1$), quantifies the benefit from leveraging pre-experimental information for $\mu_\Delta$. Looking across subfigures (i) and (ii), such height differences between methods are far greater for the unknown variance case than the known variance case. 
Comparison of SSD approaches with borrowing versus no borrowing, as visualised in subfigure (ii) of Figure \ref{fig:SScomp1}, would be more objective for illustrating the benefit. 
As mentioned, choosing $c = 3$ means  $\sigma_0^2$ would be related with $\sum p_k^2 \xi_k^2$ to a very limited extent, as if a diffuse prior had been placed on $\sigma_0^2$. Thereby, implementing no borrowing by setting $w_k = 1$, pre-experimental information would neither be leveraged through the robust prior for $\mu_\Delta$, nor through the prior for the unknown $\sigma_0^2 \sim$ Inv-Gamma($\frac{c}{2}, \frac{c\sum p_k^2 \xi_k^2}{2}$). Consequently, larger sample sizes would be found for no borrowing SSD for the unknown $\sigma_0^2$ than the known cases assuming $\sigma_0^2 = 0.35$, to retain similar properties of the posterior distribution. 
Focusing on the bars for robust weights I and II against no borrowing within subfigure (ii), saving in all the ACC, ALC and APVC sample sizes could be as much as two-thirds for configurations 1 and 2.  Such saving is attenuated in configurations 3 and 4 when historical information is divergent.
In configurations 3, the ACC (ALC) sample size obtained from the no borrowing approach is about twice the size from the proposed approach with robust weights I, specifically, 232.2 versus 116.8 (136 versus 65), respectively. We observe a small increase in sample size by using robust weights II instead of I, because slightly higher prior probabilities of incommensurability had been allocated to certain informative $N(m_k, s_k^2)$ for greater down-weighting. The trend is similar for results in configuration 4.

We then compare the proposed approach with an alternative strategy, that is, restricting the use of pre-experimental information from a single source. When the historical data are consistent (divergent) between themselves, the proposed SSD formulae lead to smaller (larger) sample sizes, as presented obviously in configuration 1 (configurations 3 and 4) for both cases of known and unknown $\sigma_0^2$. As one may perceive, such selection of a single source could be less robust than averaging over all available pre-experimental information. 
Another noteworthy finding is concerned with the comparison of the ACC and ALC sample sizes, particularly when $\sigma_0^2$ is unknown and we place a very weakly-informative prior on it (setting $c=3$). As shown in  Figure \ref{fig:SScomp1}, the ALC sample size is universally smaller than the ACC sample size for all these investigated configurations.


\begin{figure}[!ht]
\centering
\includegraphics[width=0.8\linewidth]{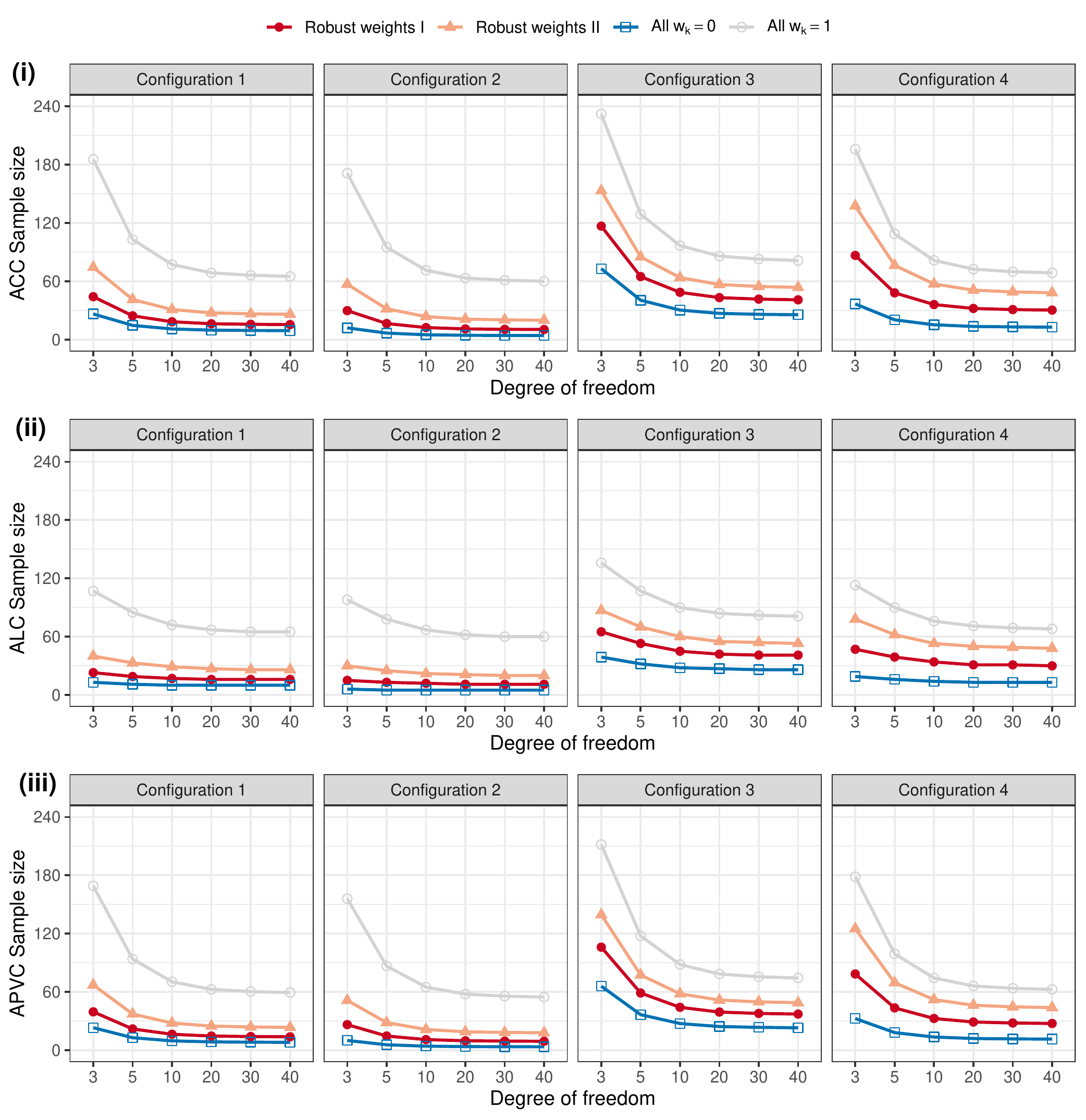}
\caption{The ACC, ALC and APVC sample sizes for the new trial, where the unknown $\sigma_0^2$ could be related to the collective prior variance by assuming the quantity $c\sum p_k^2\xi_k^2/\sigma_0^2 \sim \chi^2(c)$. The extent of borrowing for better knowledge about $\sigma_0^2$ depends on the number of degrees of freedom, $c$.}
\label{fig:SScomp2}
\end{figure}

We move on to quantify how the sample sizes would vary as $c$ changes.
Focusing on approaches using pre-experimental information from multiple sources, Figure \ref{fig:SScomp2} displays the sample sizes exclusively for cases of unknown $\sigma_0^2 \sim$ Inv-Gamma($\frac{c}{2}, \frac{c\sum p_k^2 \xi_k^2}{2}$).  We set $c = 3, 5, 10, 20, 30, 40$ and keep the target level of each SSD criterion unchanged from what we have used for Figure \ref{fig:SScomp1}.  As $c$ gets larger, the sample sizes for all approaches investigated here decrease and tend to stablise at their own lowest levels possible. This could be explained from the perspective of prior effective sample size \citep{BIOM:Neuenschwander2020}, to which variance is a key determining factor. 
Consider the prior placed on the inverse of the unknown variance that $\frac{1}{\sigma_0^2}\sim$ Gamma($\frac{c}{2}, \frac{c\sum p_k^2 \xi_k^2}{2}$), of which the mean and variance are $\frac{1}{\sum p_k^2 \xi_k^2}$ and $\frac{2}{c}\cdot\frac{1}{(\sum p_k^2 \xi_k^2)^2}$, respectively.  
As $c$ increases, the prior variance dinimishes, meaning that possible values of $\frac{1}{\sigma_0^2}$ are more concentrated around the prior mean obtained based on historical data.  
For $c\geq 20$, the ACC and ALC sample sizes are nearly identical.  
Whereas, the ACC sample size is more sensitive than the ALC to small values of $c$, e.g., when $c = 3, 5$. 
We note that the so-called `no borrowing' (by setting $w_k = 1$) should be clarified as no borrowing in terms of the parameter $\mu_\Delta$. When $c$ gets larger, it means the unknown variance $\sigma_0^2$ would be more closely tied to the prior variance based on the historical data. That is, borrowing is enabled through the variance, although not directly the parameter of inferential interest.
By fixing $w_k = 1$, historical data would not be leveraged through the robust prior for $\mu_\Delta$, but nevertheless could be used to inform the unknown $\sigma_0^2$, particularly when $c$ is sufficiently large.


\begin{figure}[!ht]
\centering
\includegraphics[width=0.8\linewidth]{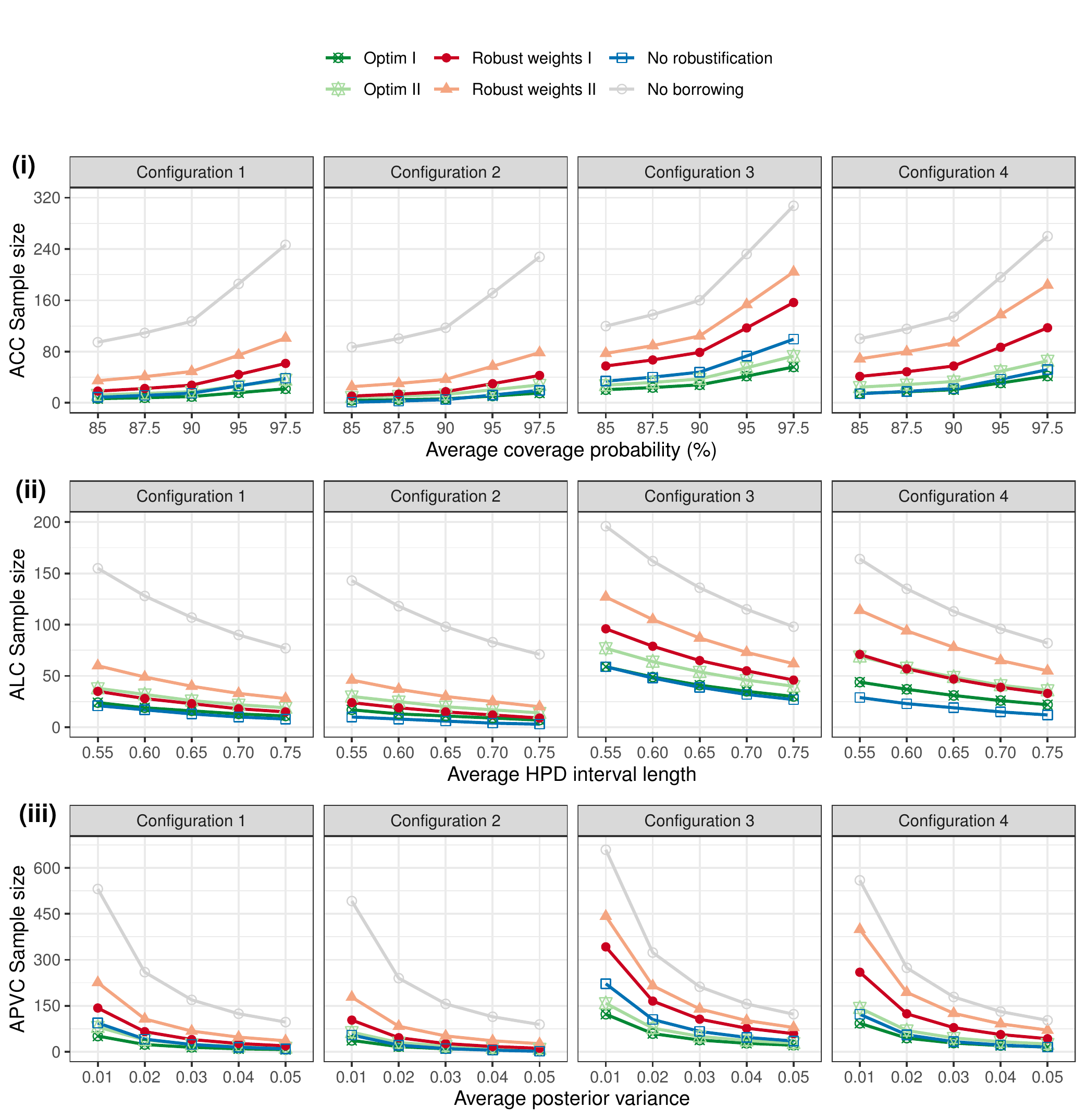}
\caption{Sample sizes required when $\sigma_0^2$ is unknown to retain desired average property of the posterior distribution. The ACC and ALC sample sizes are computed by fixing the credible interval length $\ell_0 = 0.65$ and coverage probability $1 - \alpha_0 = 95$\%, respectively.}
\label{fig:SSpar}
\end{figure}

Figure \ref{fig:SSpar} illustrates how the sample size varies, for cases of unknown $\sigma_0^2$, when targeting the average coverage probability, posterior credible length and posterior variance at different levels. Like in Figure \ref{fig:SScomp1}, these results are obtained by setting $c=3$ for the very limited use of pre-experimental information to inform $\sigma_0^2$. 
The optimal sample sizes are also plotted to show the maximal saving the proposed SSD formulae may achieve. Specifially, the optim I and II should be taken as the benchmark for formulae using robust weights I and II, respectively. 
As expected, sample sizes by robust weights I and II would always be bounded by the extremes of no robustification (all $w_k = 0$) and no borrowing (all $w_k = 1$).
Given a fixed length $\ell_0=0.65$ of the HPD interval, more ACC sample sizes would be required if increasing the desired coverage probability on average, $1-\alpha$. For example, the ACC sample size computed using robust weights I (II) rises from 78.7 to 156.5 (104.4 to 204.2) for configuration 3, had the level of $1-\alpha$ been lifted from 90\% to 97.5\%.
The displayed ALC sample sizes in subfigure (ii) ensures the coverage probability as 95\%; by relaxing the target average HPD interval length, fewer sample sizes would be needed.
Likewise, the APVC sample sizes in subfigure (iii) share this commonality of decreasing as we relax the target posterior variance.
Generating these plots would be helpful in practice for balancing between obtaining an economic sample size planning and a posterior sufficiently informative for inferences on a case-by-case basis.  For example, targeting the average length of the HPD interval with 95\% coverage probability as $\ell = 0.60$ requires the ALC sample size to be 28 for configuration 1 using robust weights I, which may not be much different from 23 yielded by the level $\ell = 0.65$.


\begin{figure}[!ht]
\centering
\includegraphics[width=0.7\linewidth]{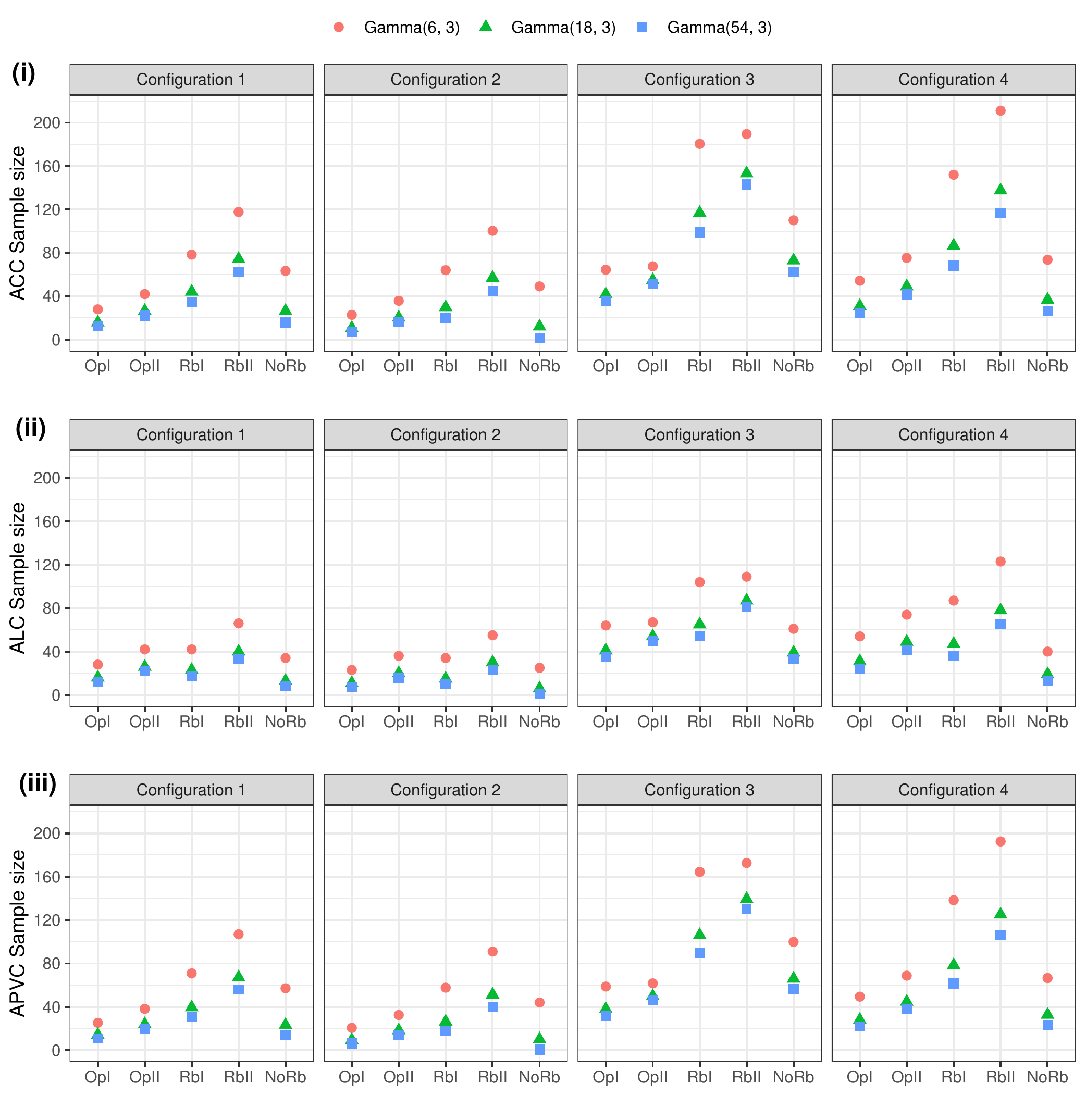}
\caption{The proposed Bayesian SSD dependent on the choice of the informative Gamma component distribution for strong borrowing. The labels at $x$-axis are short for Optim I, Optim II, Robust weights I, Robust weights II, and No robustification, respectively.}
\label{fig:GammaMix}
\end{figure}

Finally, we examine how sensitive the proposed Bayesian SSD formulae is to the Gamma mixture components. 
Since a suitable yet least informative Gamma$(a_{01},\, b_{01})$ has been chosen for down-weighting, the other component of the mixture prior, Gamma$(a_{02},\, b_{02})$, determines the maximum borrowing possible. 
Assuming unknown $\sigma_0^2$ and setting $c = 3$,
Figure \ref{fig:GammaMix} shows the Bayesian SSD under different choices of the hyperparameters, $a_{02}$ and $b_{02}$, for each criterion. As expected, a more informative Gamma$(a_{02},\, b_{02})$ yields a smaller sample size given the same set of $w_k, \, k=1, \dots, K$. The ALC sample sizes appear to have least decreasing, compared with the ACC and APVC, in this sensitivity evaluation. 
We also observe that the reduction in Bayesian sample sizes is not proportional to the improving of informativeness of Gamma$(a_{02},\, b_{02})$: setting the informative component as Gamma(18, 3) is not much different from Gamma(54, 3) for our illustrative examples. 
For practical implementation, we recommend the component Gamma distributions to be chosen for representing two extremes of very limited borrowing and complete pooling of information, when given a full prior mixture weight $w_k = 1$ and $w_k = 0$, respectively.

\section{Discussion}
\label{sec:discuss}

Planning a new experiment with a sufficient sample size necessitates the use of relevant information. 
Bayesian methods allow for the inherent uncertainty in the estimate of model parameters, as well as a formal incorporation of any expert opinion or historical data.
In this paper, we have developed Bayesian sample size formulae that use commensurate priors to leverage pre-experimental data, available from multiple sources, for the model parameter(s) of interest. The proposed approach falls within the remit of proper Bayesian approach to sample size determination, wherein the priors for design planning and data analysis remain the same.
The level of down-weighting, in the light of possible disagreement between any pre-experimental and the new experimental data, relies on the user's choice of the prior probabilities of incommensurability, $w_k, \, k = 1, \dots, K$.  
As shown in the performance evaluation, smaller values of $w_k$ imply less down-weighting, resulting in smaller sample sizes required for the new experiment. By contrast, a large value of $w_k$ for large down-weighting and thus results in large sample size.

Choosing sensible values of $w_k,\, k = 1, \dots, K,$ is crucial for practical implementation.  Following \citet{BIOS:ZhengWason2019}, we recommended these to be linked with measures of pairwise distributional discrepancy. In our illustration, the squared Hellinger distance between any two pre-experimental parameters, $\theta_k \mid \boldsymbol{y}_k$, was computed to inform the choices of $w_k$. The underlying logic is that the new experiment, at the planning stage, may be regarded as compatible with the historical experiments, and so would their data be. The levels of pairwise (in)commensurability between a pre-experimental parameter and the new experimental parameter would thus be comparable to those between the pre-experimental parameters themselves.
Nevertheless, we recognise that these prior mixture weights $w_k$ can not be correctly specified when the new experimental data are yet to be generated.
From a pragmatic perspective, the new experiment could be embedded with one or multiple interim analyses to enable mid-course modifications towards $w_k$. Each update in terms of $w_k$ tends to better reflect the genuine incommensurability \citep{BJ:Zheng2020}. This area deserves further investigation; similar topics concerning the sample size re-estimation have been discussed among others \citep{BIOM:Cui1999, BJ:Wang2007, SiM:Brutti2009, SMMR:Brakenhoff2019}.


While this paper has focused on estimating the difference between two normal means, our proposed sample size formulae are straightforward to derive for a single normal mean. 
For planning a new experiment that generates data on binomially distributed outcomes, we used logit transformation to consider the log-odds ratio as a continuous variable that could be adequately modelled by a normal distribution. It would be interesting to extend the proposed methodology for bionomial proportions in a more direct manner \citep{JRSSD:Joseph1995, BA:Mlan2008, BJ:Joseph2019}, as well as in the context of other sampling distributions. We also consider developing sample size formulae in a generalised linear regression modelling framework where potential confounding can be adjusted for.

When illustrating the application, we supposed that pre-experimental information had been made available with regard to the parameter of influential interest. In practical implementation, situations may be more complex. For instances, historical data may have been collected from experiments with very different design considerations \citep{SMMR:Zhang2019}, or recorded on a different measurement scale \citep{SMMR:Zheng2020} from what might be for the new experiment under planning. This is an area where our future research would look towards. 
As a separate note, we applied quite general criteria such as ACC and ALC to control the average coverage probability or length of the HPD interval of the posterior distribution for the parameter of influential interest throughout. 
We are aware of occasions where error rates need to be controlled, e.g., in most clinical trials for drug development undertaken in patients with non-rare diseases, as required by the regulatory agencies \citep{EMA:trials1998}. We note that our sample size formulae according to the ACC can be easily extended to give a solution, which is analogous to the frequentist hypothesis testing: rejection of the null hypothesis would be defined based on posterior interval probabilities with respect to certain magnitude of effect size  \citep{SiM:Whitehead2008}.

\section*{Acknowledgment}

JW is funded by the UK Medical Research Council (MC\rule{2mm}{0.4pt}UU\rule{2mm}{0.4pt}00002/6).  T Jaki received funding from UK Medical Research Council (MC\rule{2mm}{0.4pt}UU\rule{2mm}{0.4pt}0002/14). This report is independent research arising in part from Prof Jaki's Senior Research Fellowship (NIHR-SRF-2015-08-001) supported by the National Institute for Health Research. The views expressed in this publication are those of the authors and not necessarily those of the NHS, the National Institute for Health Research or the Department of Health and Social Care (DHSC).

\clearpage

\bibliographystyle{apalike}



\clearpage



\section*{Supplementary material}

# Web-based Supplementary Materials for:
Bayesian sample size determination using commensurate priors to leverage pre-experimental data

**by Haiyan Zheng, Thomas Jaki, James M. S. Wason**

### A. DERIVATION OF THE UNIMODAL $t$ MIXTURE DISTRIBUTION

Given the joint distribution of $\tilde{\theta}_k$ and $\nu_k$ as a mixture of Normal-Gamma distributions, it follows that

$$f(\tilde{\theta}_k \mid \theta_k) = \int f(\tilde{\theta}_k, \nu_k \mid \theta_k) \mathrm{d}\nu_k = \int f(\tilde{\theta}_k \mid \theta_k, \nu_k) g(\nu_k) \mathrm{d}\nu_k$$

$$= w_k \frac{b_{01}^{a_{01}}}{\sqrt{2\pi}\Gamma(a_{01})} \int_0^\infty \nu_k^{a_{01}-\frac{1}{2}} \exp\left(-\frac{2b_{01}\nu_k + \nu_k(\tilde{\theta}_k - \theta_k)^2}{2}\right) \mathrm{d}\nu_k +$$

$$(1-w_k) \frac{b_{02}^{a_{02}}}{\sqrt{2\pi}\Gamma(a_{02})} \int_0^\infty \nu_k^{a_{02}-\frac{1}{2}} \exp\left(-\frac{2b_{02}\nu_k + \nu_k(\tilde{\theta}_k - \theta_k)^2}{2}\right) \mathrm{d}\nu_k \quad \text{(S1)}$$

$$= w_k \frac{b_{01}^{a_{01}}}{\sqrt{2\pi}\Gamma(a_{01})} \int_0^\infty \nu_k^{a_{01}-\frac{1}{2}} \exp\left(-\frac{\nu_k((\tilde{\theta}_k - \theta_k)^2 + 2b_{01})}{2}\right) \mathrm{d}\nu_k +$$

$$(1-w_k) \frac{b_{02}^{a_{02}}}{\sqrt{2\pi}\Gamma(a_{02})} \int_0^\infty \nu_k^{a_{02}-\frac{1}{2}} \exp\left(-\frac{\nu_k((\tilde{\theta}_k - \theta_k)^2 + 2b_{02})}{2}\right) \mathrm{d}\nu_k.$$

Let $z_A = \frac{\nu_k}{2} \cdot A$, where $A = (\tilde{\theta}_k - \theta_k)^2 + 2b_{01}$, and let $z_B = \frac{\nu_k}{2} \cdot B$, where $B = (\tilde{\theta}_k - \theta_k)^2 + 2b_{02}$. We can re-write (S1) as follows.

$$f(\tilde{\theta}_k \mid \theta_k) = \int f(\tilde{\theta}_k, \nu_k \mid \theta_k) \mathrm{d}\nu_k$$

$$= w_k \frac{b_{01}^{a_{01}}}{\sqrt{2\pi}\Gamma(a_{01})} \left(\frac{2}{A}\right)^{a_{01}+\frac{1}{2}} \int z_A^{a_{01}-\frac{1}{2}} \exp(-z_A) \mathrm{d}z_A + \quad \text{(S2)}$$

$$(1-w_k) \frac{b_{02}^{a_{02}}}{\sqrt{2\pi}\Gamma(a_{02})} \left(\frac{2}{B}\right)^{a_{02}+\frac{1}{2}} \int z_B^{a_{02}-\frac{1}{2}} \exp(-z_B) \mathrm{d}z_B,$$

where $\int z_A^{a_{01}-\frac{1}{2}} \exp(-z_A) \mathrm{d}z_A$ and $\int z_B^{a_{02}-\frac{1}{2}} \exp(-z_B) \mathrm{d}z_B$ are Gamma intergrals. Therefore,

$$f(\tilde{\theta}_k \mid \theta_k) \propto w_k b_{01}^{a_{01}} \left(\frac{2}{A}\right)^{a_{01}+\frac{1}{2}} + (1-w_k) b_{02}^{a_{02}} \left(\frac{2}{B}\right)^{a_{02}+\frac{1}{2}}$$

$$\propto w_k \left(\frac{1}{b_{01}}\right)^{-a_{01}} \left(\frac{A}{2}\right)^{-a_{01}-\frac{1}{2}} + (1-w_k)\left(\frac{1}{b_{02}}\right)^{-a_{02}} \left(\frac{B}{2}\right)^{-a_{02}-\frac{1}{2}}$$

$$\propto w_k \left(\frac{1}{b_{01}}\right)^{-a_{01}-\frac{1}{2}} \left(\frac{(\tilde{\theta}_k - \theta_k)^2 + 2b_{01}}{2}\right)^{-a_{01}-\frac{1}{2}} + \quad \text{(S3)}$$

$$(1-w_k)\left(\frac{1}{b_{02}}\right)^{-a_{02}-\frac{1}{2}} \left(\frac{(\tilde{\theta}_k - \theta_k)^2 + 2b_{02}}{2}\right)^{-a_{02}-\frac{1}{2}}$$

$$\propto w_k \left(\frac{(\tilde{\theta}_k - \theta_k)^2}{2b_{01}} + 1\right)^{-\frac{2a_{01}+1}{2}} + (1-w_k)\left(\frac{(\tilde{\theta}_k - \theta_k)^2}{2b_{02}} + 1\right)^{-\frac{2a_{02}+1}{2}},$$

which demonstrates that the predictive prior distribution for each $\tilde{\theta}_k$, given a historical parameter $\theta_k$, is a mixture of non-standardised $t$ distributions. In particular, the location parameters are *identical* to $\theta_k$, while the scale parameters are $\frac{b_{01}}{a_{01}}$ and $\frac{b_{02}}{a_{02}}$, respectively. In other words, the two mixture components correspond to $\frac{\tilde{\theta}_k - \theta_k}{\sqrt{b_{01}/a_{01}}}$ and $\frac{\tilde{\theta}_k - \theta_k}{\sqrt{b_{02}/a_{02}}}$ that follow classic Student's $t$ distributions with $2a_{01}$ and $2a_{02}$ degrees of freedom, respectively.

## B. MOMENTS AND OVERALL VARIANCE OF THE $t$ MIXTURE DISTRIBUTION

For a finite mixture distribution denoted by $f(x) = \sum_i w_i h_i(x)$, its $\ell$-th moment about zero exists and follows that

$$\mu^{(\ell)} = \mathbb{E}_f(x^{(\ell)}) = \sum_i w_i \mathbb{E}_{h_i}(x^{(\ell)}) = \sum_i w_i \mu_i^{(\ell)}, \tag{S4}$$

where $\mu_i^{(\ell)}$ is each $\ell$-th moment of the mixture component $h_i(x)$. In our context, the constant prior mixture weights $w_1 = w_k$ and $w_2 = 1 - w_k$; while the components $h_i(\tilde{\theta}_k|\theta_k)$, $i = 1, 2$, are the non-standardised Student's $t$ distributions, both centred at $\theta_k$. We then know that $\mu_1^{(1)} = \mu_2^{(1)} = \theta_k$. We further constrain that $a_{01}, a_{02} > 1$ so that variances, denoted by $\sigma_i^2$, of $h_i(\tilde{\theta}_k|\theta_k)$ exist, as $\frac{b_{01}}{a_{01}-1}$ and $\frac{b_{02}}{a_{02}-1}$, respectively. Following Equation (S4), the overall variance for the mixture distribution can be computed as

$$\begin{aligned}
\text{Var}(\tilde{\theta}_k|\theta_k) &= \mu^{(2)} - [\mu^{(1)}]^2 \\
&= \sum_i w_i \mu_i^{(2)} - \left[\sum_i w_i \mu_i^{(1)}\right]^2 \\
&= \sum_i w_i(\sigma_i^2 + [\mu_i^{(1)}]^2) - \left[\sum_i w_i \mu_i^{(1)}\right]^2 \\
&= \sum_i w_i \sigma_i^2 + \sum_i w_i [\mu_i^{(1)}]^2 - \left[\sum_i w_i \mu_i^{(1)}\right]^2 \\
&= w_k \frac{b_{01}}{a_{01}} \cdot \frac{2a_{01}}{2a_{01} - 2} + (1 - w_k)\frac{b_{02}}{a_{02}} \cdot \frac{2a_{02}}{2a_{02} - 2} + w_k \theta_k^2 + (1 - w_k)\theta_k^2 - [w_k \theta_k + (1 - w_k)\theta_k]^2 \\
&= \frac{w_k b_{01}}{a_{01} - 1} + \frac{(1 - w_k) b_{02}}{a_{02} - 1}.
\end{aligned}$$

## C. ON THE GOODNESS OF THE NORMAL APPROXIMATION

As illustrated, $\tilde{\theta}_k$ follows a non-standard $t$ mixture distribution that has its two components both centred at $\theta_k$. We approximate it by a normal distribution $N\left(\theta_k, \frac{w_k b_{01}}{a_{01}-1} + \frac{(1-w_k)b_{02}}{a_{02}-1}\right)$, of which the variance was set equal to the overall variance of the unimodal $t$ mixture prior.

This approximation is considered to be appropriate for the comparable coverage probabilities of a given credible interval, defined as $(\theta_k - r, \theta_k + r)$. Precisely, with the unimodal $t$ mixture

distribution, the coverage probability can be given by

$$\mathbb{P}_t(|\tilde{\theta}_k - \theta_k| \leq r) = w_k \times \int_{\theta_k-r}^{\theta_k+r} h_1(\tilde{\theta}_k)d\tilde{\theta}_k + (1-w_k) \times \int_{\theta_k-r}^{\theta_k+r} h_2(\tilde{\theta}_k)d\tilde{\theta}_k$$

$$= \frac{w_k\Gamma(\frac{2a_{01}+1}{2})}{\Gamma(a_{01})\sqrt{2\pi b_{01}}} \int_{-r}^{r} \left(1 + \frac{(\tilde{\theta}_k - \theta_k)^2}{2b_{01}}\right)^{-\frac{2a_{01}+1}{2}} d(\tilde{\theta}_k - \theta_k) + \quad (S5)$$

$$\frac{(1-w_k)\Gamma(\frac{2a_{02}+1}{2})}{\Gamma(a_{02})\sqrt{2\pi b_{02}}} \int_{-r}^{r} \left(1 + \frac{(\tilde{\theta}_k - \theta_k)^2}{2b_{02}}\right)^{-\frac{2a_{02}+1}{2}} d(\tilde{\theta}_k - \theta_k).$$

The coverage probability of $(\theta_k - r, \theta_k + r)$ given the normal approximation has a closed-form expression, and we further let it be of size $1 - \alpha$:

$$\mathbb{P}_G(|\tilde{\theta}_k - \theta_k| \leq r) = \Phi\left(\frac{r}{\sqrt{\frac{w_k b_{01}}{a_{01}-1} + \frac{(1-w_k)b_{02}}{a_{02}-1}}}\right) = 1 - \alpha.$$

We may then solve this by giving $r = z_{\alpha/2}\sqrt{\frac{w_k b_{01}}{a_{01}-1} + \frac{(1-w_k)b_{02}}{a_{02}-1}}$, where $z_{\alpha/2}$ denotes the upper $(\alpha/2)$-th quantile of the standard normal distribution, i.e., $\Phi^{-1}(1 - \alpha/2)$. Substituting this to (S5), we can derive the exact coverage probability of the interval in the non-standardised $t$ mixture distribution, which is approximately $1 - \alpha$.

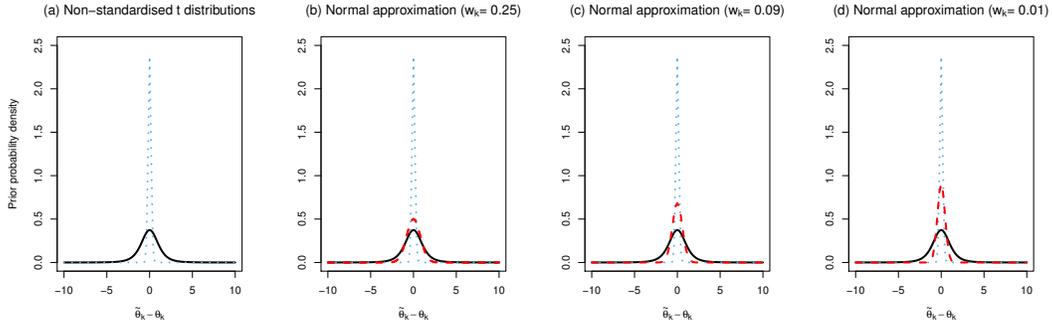

Figure S1: The $t$ mixture distribution for $\tilde{\theta}_k - \theta_k$, with the components depicted using black solid and blue dotted curves. Its normal approximations, given by $\tilde{\theta}_k - \theta_k \sim N\left(0, \frac{w_k b_{01}}{a_{01}-1} + \frac{(1-w_k)b_{02}}{a_{02}-1}\right)$, are visualised using the red dashed curves in subfigures (b) – (d), under various choices of $w_k$.

For illustration, we specify that $\nu_k \sim w_k\text{Gamma}(2, 2) + (1 - w_k)\text{Gamma}(18, 3)$. Thus, $\tilde{\theta}_k - \theta_k$ follows a non-standardised $t$ mixture distribution centered at 0, which may be approximated by a $N\left(0, \frac{w_k b_{01}}{a_{01}-1} + \frac{(1-w_k)b_{02}}{a_{02}-1}\right)$. Figure S1 displays some example density curves for $\tilde{\theta}_k - \theta_k$, with the component $t$ distributions visualised in each subfigure. In subfigures (b) – (d), we overlie the density curve of the normal approximation, given prior mixture weights $w_k = 0.25, 0.09, 0.01$. As we observe, a smaller value of $w_k$ tends to distort the approximated normal density curve further towards the more informative component $t$ distribution.

Based on the normal approximation and setting $w_k = 0.25, 0.09, 0.01$, a 95% credible interval being symmetric about $\theta_k$ can be yielded by $r = 1.559, 1.144, 0.865$, respectively. Substituting these values of $r$ to (S5), we compute the exact coverage probabilities based on the original $t$ mixture distribution as 95.1%, 96.4%, and 95.5%, which are all close to 95% as expected. Finally, we note that the component distribution Gamma(2, 2) is quite extreme; with larger $a_{01}$ and $b_{01}$, the approximation tends to be even better.

## D. PREDICTIVE PRIOR THAT LEVERAGES MULTI-SOURCE PRE-EXPERIMENTAL DATA

We start from the normal approximation to the non-standardised $t$ mixture distribution:

$$\tilde{\theta}_k \mid \theta_k \dot\sim N\left(\theta_k, \frac{w_k b_{01}}{a_{01}-1} + \frac{(1-w_k)b_{02}}{a_{02}-1}\right),$$

where $\theta_k$ has a (posterior) distribution $\theta_k|y_k \sim N(m_k, s_k^2)$, or this distribution is characterised through elicitation of expert prior opinion on the difference in means. We then write

$$\tilde{\theta}_k = \theta_k + e,$$

where, independently to $\theta_k$, we know $e \sim N\left(0, \frac{w_k b_{01}}{a_{01}-1} + \frac{(1-w_k)b_{02}}{a_{02}-1}\right)$. Now $\tilde{\theta}_k$ is just the sum of two independent normal random variables, which is fully determined by mean

$$\lambda_k = \mathbb{E}(\tilde{\theta}_k) = \mathbb{E}(\theta_k) + \mathbb{E}(e) = m_k,$$

and variance

$$\xi_k^2 = \text{Var}(\tilde{\theta}_k) = \text{Var}(\theta_k) + \text{Var}(e) = s_k^2 + \frac{w_k b_{01}}{a_{01}-1} + \frac{(1-w_k)b_{02}}{a_{02}-1}.$$

## E. POSTERIOR FOR THE DIFFERENCE IN MEANS WHEN THE VARIANCE IS UNKNOWN

We suppose that the unknown mean $\mu_\Delta$ has a normal prior distribution specified based on $K$ pre-experimental datasets, $N(\sum p_k \lambda_k, \sum p_k^2 \xi_k^2)$, and the unknown variance $\sigma_0^2$ has an Inv-Gamma$(\frac{c}{2}, \frac{c\sum p_k^2 \xi_k^2}{2})$ prior distribution. The quantity $c\sum p_k^2 \xi_k^2 / \sigma_0^2$ is then a $\chi^2(c)$ distribution. The posterior for $\mu_\Delta$ can thus be given by

$$\begin{aligned}
f_p(\mu_\Delta \mid y_1, \ldots, y_K, y_{K+1}) &= \int \pi_p(\mu_\Delta, \sigma_0^2 \mid y_1, \ldots, y_K, y_{K+1}) g(\sigma_0^2) d\sigma_0^2 \\
&\propto \int \mathcal{L}(\bar{x}_\Delta \mid \mu_\Delta, \sigma_0^2) \pi(\mu_\Delta \mid y_1, \ldots, y_K) g(\sigma_0^2) d\sigma_0^2 \\
&\propto \int \frac{1}{\sigma_0} \exp\left(-\frac{(\bar{x}_\Delta - \mu_\Delta)^2}{2\sigma_0^2 \left(\frac{1}{n_A} + \frac{1}{n_B}\right)}\right) \cdot \exp\left(-\frac{(\mu_\Delta - \sum p_k \lambda_k)^2}{2 \sum p_k^2 \xi_k^2}\right) \cdot \frac{1}{(\sigma_0^2)^{\frac{c}{2}+1}} \exp\left(-\frac{c\sum p_k^2 \xi_k^2}{2\sigma_0^2}\right) d\sigma_0^2 \\
&\propto \exp\left(-\frac{(\mu_\Delta - \sum p_k \lambda_k)^2}{2 \sum p_k^2 \xi_k^2}\right) \int \frac{1}{(\sigma_0^2)^{\frac{c+1}{2}+1}} \exp\left(-\frac{1}{2\sigma_0^2}\left[\frac{(\bar{x}_\Delta - \mu_\Delta)^2}{\frac{1}{n_A} + \frac{1}{n_B}} + c\sum p_k^2 \xi_k^2\right]\right) d\sigma_0^2 \\
&\propto \exp\left(-\frac{(\mu_\Delta - \sum p_k \lambda_k)^2}{2 \sum p_k^2 \xi_k^2}\right) \left[\frac{(\bar{x}_\Delta - \mu_\Delta)^2}{\frac{1}{n_A} + \frac{1}{n_B}} + c\sum p_k^2 \xi_k^2\right]^{-\frac{c+1}{2}}.
\end{aligned}$$
(S6)

The last step in (S6) follows, since the integrant is the kernel of the density for an Inv-Chi-square distribution with $c$ degrees of freedom, multiplied by $\left[\frac{(\bar{x}_\Delta - \mu_\Delta)^2}{\frac{1}{n_A} + \frac{1}{n_B}} + c\sum p_k^2 \xi_k^2\right]$. We then divide both terms in the square brackets by $c\sum p_k^2 \xi_k^2$ and obtain

$$f_p(\mu_\Delta \mid y_1, \ldots, y_K, y_{K+1}) \propto \exp\left(-\frac{(\mu_\Delta - \sum p_k \lambda_k)^2}{2 \sum p_k^2 \xi_k^2}\right) \left[1 + \frac{(\mu_\Delta - \bar{x}_\Delta)^2}{c\left(\frac{1}{n_A} + \frac{1}{n_B}\right) \sum p_k^2 \xi_k^2}\right]^{-\frac{c+1}{2}}, \quad (S7)$$

which features the product of kernels of normal and non-standardised $t$ distributions, with the latter of $c$ degrees of freedom, shifted by $\bar{x}_\Delta$ and scaled by $\left(\frac{1}{n_A} + \frac{1}{n_B}\right) \sum p_k^2 \xi_k^2$. Therefore, the marginal posterior relies on the distribution of $\bar{x}_\Delta$.

We can easily work out that $\bar{x}_\Delta \mid \mu_\Delta$ has a $t$ distribution, as

$$f(\bar{x}_\Delta \mid \mu_\Delta) = \int \mathcal{L}(\bar{x}_\Delta \mid \mu_\Delta, \sigma_0^2) g(\sigma_0^2) d\sigma_0^2 \propto \left[1 + \frac{1}{c} \cdot \frac{(\bar{x}_\Delta - \mu_\Delta)^2}{\left(\frac{1}{n_A} + \frac{1}{n_B}\right) \sum p_k^2 \xi_k^2}\right]^{-\frac{c+1}{2}}.$$

Likewise, we relate it with the normal kernel conditional on unknown $\sigma_0^2$ so that

$$f(\bar{x}_\Delta \mid \mu_\Delta, \sigma_0^2) \propto \frac{1}{\sigma_0} \exp\left(\frac{(\bar{x}_\Delta - \mu_\Delta)^2}{2\left(\frac{1}{n_A} + \frac{1}{n_B}\right)\sigma_0^2}\right).$$

Placing the normal prior distribution for $\mu_\Delta$, specified based on pre-experimental data, we obtain the distribution for $\bar{x}_\Delta$ unconditional on $\mu_\Delta$ as

$$\bar{x}_\Delta \mid \sigma_0^2, \mathbf{y}_1, \ldots, \mathbf{y}_K \sim N\left(\sum p_k \lambda_k, \left(\frac{1}{n_A} + \frac{1}{n_B}\right)\sigma_0^2 + \sum_k p_k^2 \xi_k^2\right),$$

which depends on the distribution for $\sigma_0^2$.

## F. PAIRWISE COMMENSURABILITY OF THE HYPOTHETICAL HISTORICAL DATA

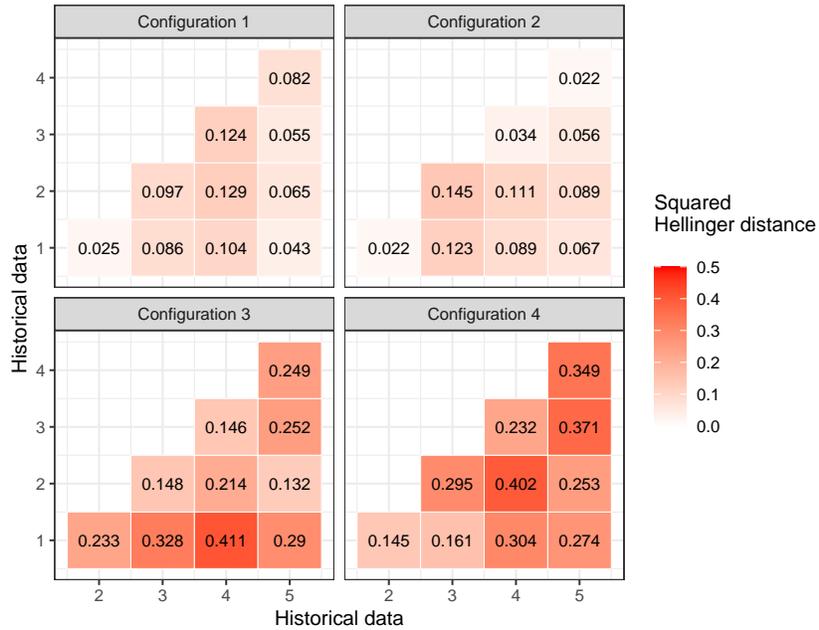

Figure S2: Pairwise distributional discrepancy between any two $\theta_k \mid \mathbf{y}_k \sim N(m_k, s_k^2)$.

We compute the squared Hellinger distance [1] for all pairs of the individually specified prior opinion on the log-odds ratio, $\theta_k \mid \mathbf{y}_k$, with each represented in a $N(m_k, s_k^2)$ distribution. The

5

squared Hellinger distance between two normal distributions, say, $N(m_k, s_k^2)$ and $N(m_{k'}, s_{k'}^2)$, has a closed-form expression:

$$H^2(\theta_k \mid \boldsymbol{y}_k, \theta_{k'} \mid \boldsymbol{y}_{k'}) = 1 - \sqrt{\frac{2s_k^2 s_{k'}^2}{s_k^2 + s_{k'}^2}} \exp\left(-\frac{(m_k - m_{k'})^2}{4(s_k^2 + s_{k'}^2)}\right).$$

Figure S2 visualised the pairwise distances for hypothetical pre-experimental data. This could be useful to determine the prior probabilities of incommensurability, $w_k$, $k = 1, \ldots, K$, for applying the proposed Bayesian sample size formulae.

**References**

1. Dey, D. K. and Birmiwal, L. R. (1994). Robust Bayesian analysis using divergence measures. *Statistics & Probability Letters* **20**(4), 287 – 294.



\end{document}